\documentstyle[11pt]{article}

\newcommand{\noi}{\vspace{12pt}\noindent}

\newcommand{\ie}{{${ i.e.\ }$}}

\newcommand{\beq}{\begin{equation}}
\newcommand{\eeq}{\end{equation}}
\newcommand{\bea}{\begin{eqnarray}}
\newcommand{\eea}{\end{eqnarray}}
\newcommand{\lpart}{\raise.3ex\hbox{$\stackrel{\leftarrow}{\partial}$}}
\newcommand{\rpart}{\raise.3ex\hbox{$\stackrel{\rightarrow}{\partial}$}}
\newcommand{\ldr}{\raise.3ex\hbox{$\stackrel{\leftarrow}{\delta^r}$}}
\newcommand{\deder}[1]{{ 
 {\stackrel{\raise.1ex\hbox{$\leftarrow$}}{\delta^r}   } 
\over {   \delta {#1}}  }}
\newcommand{\dedel}[1]{{ 
 {\stackrel{\lower.3ex \hbox{$\rightarrow$}}{\delta^l}   }
 \over {   \delta {#1}}  }}
\newcommand{\dede}[1]{{   {\delta} \over {\delta {#1}}  }}
\newcommand{\papar}[1]{{ 
 {\stackrel{\raise.1ex\hbox{$\leftarrow$}}{\partial^r}   } 
\over {   \partial {#1}}  }}
\newcommand{\papal}[1]{{ 
 {\stackrel{\lower.3ex \hbox{$\rightarrow$}}{\partial^l}   }
 \over {   \partial {#1}}  }}
\newcommand{\zzzz}{z}
\newcommand{\Gammm}{\Gamma}
\newcommand{\hEo}{\hat{E}_{0}}
\newcommand{\Eo}{E_{0}}
\newcommand{\EE}{E}
\newcommand{\vol}{\Omega_{\rm vol}}
\newcommand{\Gameff}{\Gamma_{\rm eff}}

\newcommand{\hf}{{\scriptstyle{1 \over 2}}}
\newcommand{\Hf}{{1 \over 2}}

\newcommand{\Ih}{{i \over \hbar}}

\newcommand{\e}[1]{{(\ref{#1})}}
\newcommand{\eq}[1]{{eq.\ (\ref{#1})}}
\newcommand{\eqs}[2]{{eqs.\ (\ref{#1}) and (\ref{#2})}}
\newcommand{\mb}[1]{{\mbox{${#1}$}}}
\begin{document}
\thispagestyle{empty}
\vspace{3cm}
\title{\Large{\bf Superfield Quantization}}
\vspace{2cm}
\author{{\sc I.A.~Batalin}\\I.E.~Tamm Theory Division\\
P.N. Lebedev Physics Institute\\Russian Academy of Sciences\\
53 Leniniski Prospect\\Moscow 117924\\Russia\\
~\\~{\sc K.~Bering}\thanks{Address after Sept. 1st, 1997: Center for 
Theoretical Physics, MIT, Cambridge MA 02139, USA.}
\\Institute of Theoretical Physics\\
Uppsala University\\P.O. Box 803\\S-751 08 Uppsala\\Sweden\\~\\and\\~\\
{\sc P.H.~Damgaard}\\The Niels Bohr Institute\\Blegdamsvej 17\\
DK-2100 Copenhagen\\Denmark}
\maketitle
\begin{abstract}
We present a superfield formulation of the quantization program for
theories with first class constraints. An exact operator formulation
is given, and we show how to set up a phase-space path integral
entirely in terms of superfields. BRST transformations and canonical 
transformations enter on equal footing, and they allow us to establish
a superspace analog of the BFV theorem. We also present a formal derivation
of the Lagrangian superfield analogue of the field-antifield 
formalism, by an integration over half of the phase-space variables.  
\end{abstract}
\vspace{1.8cm}
\begin{flushleft}
NBI-HE-97-38\\UUITP-10/97\\hep-th/9708140
\end{flushleft}

\vfill
\newpage

\setcounter{equation}{0}
\section{Introduction}

\noi
When quantizing gauge theories of the most general kind, an absolutely
fundamental r\^{o}le is played by a rigid symmetry that transforms bosons
into fermions, and vice versa. Ever since its discovery,
be it in the Hamiltonian \cite{BFV,BF} or Lagrangian \cite{BRST}
formulation, it has been a dream to make this symmetry manifest by means 
of a superfield formulation. Some of the first preliminary steps in this
direction were taken in ref.\ \cite{FOS}, and there has since been an
enormous amount of literature on the subject (see, $e.g.$, ref.\
\cite{HSV} for a more recent approach, and references therein). 
Almost all of these attempts have been restricted to the Lagrangian
formalism, and it is probably fair to say that a completely general
formulation has not been found yet.

\noi 
In view of this, we find it appropriate to reconsider the problem from
a completely new direction. As is usually the case, it is instructive 
to go back to the basic starting point, which is the operator formalism.
There the essential ingredients are the equal-time canonical commutation 
relations and the Heisenberg equations of motion, which describe the
time evolution of the system. The needed superspace will therefore be
two-dimensional, consisting of just $t$, ordinary time, and a new
Grassmann-odd direction, which we denote by $\theta$. All coordinates
and momenta, and all operators involving them, must then be generalized to 
superfield operators. Because of the Grassmann-odd nature of the
additional coordinate, any object $A(t,\theta)$ defined in terms of these 
basic superfields will have a formal Taylor expansion in $\theta$ that 
truncates after just one term:
\beq
A(t,\theta) ~=~ A_{0}(t) + \theta A_{1}(t) ~.
\eeq
Therefore $A(t,\theta)$ will have the same statistics as $A_0(t)$. 
While the superspace of this construction is inherently
two-dimensional, it in an obvious manner becomes extended to a
$(d+1)$-dimensional superspace of coordinates $(x^{\mu},\theta)$ when
considered in the context of a Lorentz covariant quantum field theory
in $d$ dimensions.

\noi
Obtaining a superfield formulation of the quantization program has the
obvious advantage that the fermion-boson symmetry\footnote{{}From now
on we for simplicity denote this ``BRST symmetry'', even when
referring to its Hamiltonian counterpart.} between all required
fields, ghosts, ghosts-for-ghosts, Lagrangian multipliers, etc.,
becomes manifest. All standard superspace techniques are then applicable
to the analysis of the perturbative expansion. The BRST symmetry in
particular will be kept in a manifest manner at all stages of the
computation. At a purely conceptual level it is pleasing to see that, as has
long been suspected, the BRST symmetry can be understood in terms of
an extension of space-time to include an additional fermionic direction.
This puts the quantization 
program for dynamical systems with first-class constraints into a new 
geometrical framework.

\noi
We shall not here be concerned with the application of the superfield
formalism to specific (field) theories, but shall rather seek a completely
general framework, independent of the underlying Hamiltonian dynamics.
Our initial input shall thus be the already proven existence of suitable
BRST operators for the dynamics governed by first-class constraints. This
entails the extension of the symplectic phase space to a possibly huge
set of additional ghost fields and Lagrangian multipliers, all of which
has been completely and rigorously established \cite{BFV,BF0,BF}. 

\noi
Once the required set of fields has been introduced, the essential three
ingredients of the quantization program in the Hamiltonian formalism
are: 1) The bosonic Hamiltonian operator $\hat{H}$, ~2) The fermionic
BRST operator $\hat{\Omega}$, and ~3) A fermionic gauge-fixing function
$\hat{\Psi}_{1}$. In our sought-for superfield formulation, these three
objects must be grouped into suitable combinations. It turns out to be
most natural to link together the Hamiltonian and the BRST operator into
one single (fermionic) object \mb{\hat{Q} = \hat{\Omega} + \theta\hat{H}},
which is nilpotent. The fermionic object $\hat{\Psi}_{1}$ is thus left
over, without a bosonic superfield partner. As might have been expected
from the operator quantization program, this is due to the fact that
the one other fundamental freedom of the theory, that of canonical 
transformations on the phase space, has not been taken into account.  
Just as the
(bosonic) Hamiltonian is linked together with the (fermionic) generator
of BRST transformations, so the (fermionic) gauge-fixing function
$\hat{\Psi}_{1}$ is linked with a (bosonic) generator of canonical 
transformations. All four objects enter in a tightly-knit way, and it
should then come as no surprise that also proofs of gauge independence
in this superfield formulation involve the use of both types of 
transformations. 

\noi
Our paper is organized as follows. In the next section we introduce
the required superspace derivatives, and postulate the superfield operator
equations of motion. Consistency conditions are found to be satisfied,
and the resulting dynamics reduces, in the subsector of original
variables to that of the original Heisenberg equations of motion. We
proceed to derive, in superfield language, the gauge independence of
physical matrix elements, using the exact operator formulation. In
Section \ref{secpathint} we propose a superfield phase-space path integral, 
and demonstrate
that it satisfies all the criteria one must require for consistency with
original path integral formulation. In particular, as one of the most
important steps of this section, we establish a superfield version
of the BFV Theorem. In Section \ref{secantifield} 
we show how to formally derive an
analogous superfield Lagrangian path integral by the formal integration
over half of the phase-space variables. Introducing suitable sources
for BRST-trasnforms in the Hamiltonian language, we are naturally led to
a superfield formulation of the field-antifield formalism. In particular,
we can formally show that the superfield action must satisfy a superfield
quantum Master Equation based on a nilpotent superfield $\Delta$-operator.
Section \ref{secconclus} contains our conclusions. 
In Appendix A we outline an
alternative proof of the superfield BFV Theorem, here using a more
symmetric set of transformations in the path integral. We give our
conventions, and some useful superfield identities, in Appendix B.

\setcounter{equation}{0}
\section{The Operator Formalism}
\label{secopform}

\noi
Let us begin by introducing some notation. We consider a dynamical system 
with first-class constraints (generators of gauge symmetries), and of
phase-space dimension $2N$. The phase space variables are denoted
collectively by \mb{\zzzz_0^A(t)}. They have general Grassmann parities 
\mb{\epsilon(\zzzz_0^A) \equiv \epsilon_A}. 
We distinguish usual $c$-numbers from their associated operator 
counterparts by means of hats. Thus, for example, \mb{\hat{\zzzz}_0^A(t)} 
are the operators corresponding to \mb{\zzzz_0^A(t)}. 
Grassmann parities of operators are of course inherited from their 
$c$-number ancestors. The supercommutator of two operators is defined
as usual:
\beq
[\hat{A},\hat{B}] ~\equiv~ \hat{A}\hat{B} - (-1)^{\epsilon(\hat{A})
\epsilon(\hat{B})} \hat{B}\hat{A}  ~.
\eeq

\noi
We begin by making a superfield extension of the phase space variables.
In operator form,
\beq
\hat{\zzzz}^{A}(t,\theta) ~\equiv~ \hat{\zzzz}_{0}^{A}(t) + \theta
\hat{\zzzz}_{1}^{A}(t) ~. \label{gammasuper}
\eeq
Obviously, the superpartner 
\mb{\epsilon(\zzzz_{1}^{A}) = \epsilon_{A} +1}
has opposite statistic. The superfield \mb{\hat{\zzzz}^{A}(t,\theta)} 
is uniquely determined by a first order equation of motion (see Section 
\ref{seceqofmot}) and initial data at time $t_i$ and \mb{\theta_{i}=0}.

\noi
Quantization is imposed by
equal-$t$-equal-$\theta$ canonical commutator relations 
\beq
[\hat{\zzzz}^A(t,\theta), \hat{\zzzz}^B(t,\theta)] 
~=~ i\hbar\hat{\omega}^{AB}(t,\theta)~, \label{cqr} 
\eeq
In order to maintain this quantization relation for all pairs $(t,\theta)$,
the  \mb{\hat{\omega}^{AB}(t,\theta)}
should commute with the evolution operator 
in the $t$ and the $\theta$ direction. 
We shall guarantee this by imposing the simplifying ansatz
\beq
 \hat{\omega}^{AB}(t,\theta)~=~  \omega^{AB} \hat{{\bf 1}}~,
\label{quantansatz}
\eeq
for some {\em constant} invertible symplectic metric $\omega^{AB}$.
This commutes trivially with any evolution.
As a result {\em quantization} (\ie the process of reducing commutators
with the help of \eq{cqr}) and {\em translation} in the 
$t$ and $\theta$ directions commute. For an arbitrary operator 
\mb{\hat{A} (\hat{\zzzz}^A(t,\theta),t,\theta)} 
we define $(t',\theta')$-translations $\tau_{(t',\theta')}$ 
by
\beq
( \tau_{(t',\theta')}\hat{A} )( \hat{\zzzz}^A(t,\theta),t,\theta)
~\equiv~\hat{A}( \hat{\zzzz}^A(t-t',\theta-\theta'),t-t',\theta-\theta')~.
\eeq
We therefore have 
\beq
  \tau_{(t',\theta')} [\hat{A},\hat{B}]
~=~ [\tau_{(t',\theta')}\hat{A}  ,  \tau_{(t',\theta')}\hat{B}] ~,
\label{tauquantcom}
\eeq
where it is understood that all commutators are replaced
with the l.h.s.\ of the canonical quantization relations \e{cqr}. 
Using the principle that
quantization and translation should commute we see that the
quantization relation \eq{cqr} (with ansatz \e{quantansatz}) 
is equivalent to the ordinary equal-time relation:
\beq
[\hat{\zzzz}_{0}^A(t), \hat{\zzzz}_{0}^B(t)] 
~=~ i\hbar\omega^{AB} \hat{{\bf 1}}~. \label{cqr0} 
\eeq

\noi
We now define two superspace derivatives by
\beq
D ~\equiv~ \frac{d}{d\theta} + \theta \frac{d}{dt} ~,~~~~~~~~~~~~
\bar{D} ~\equiv~ \frac{d}{d\theta} - \theta \frac{d}{dt} ~,
\label{d}
\eeq
and, since we shall also be considering explicit differentiation,
\beq
{\cal D} ~\equiv~ \frac{\partial}{\partial\theta}
 + \theta  \frac{\partial}{\partial t} ~,~~~~~~~~~~~
\bar{{\cal D}} ~\equiv~ \frac{\partial}{\partial\theta} 
- \theta  \frac{\partial}{\partial t}~.
\eeq

\noi
Note that, as expected, $D$ and $\bar{D}$ both act like ``square roots'' 
of the time derivative:
\beq
D^2 ~=~ \frac{d}{dt}~=~- \bar{D}^2~,~~~~~~~~~~
[D,\bar{D}]~=~0 ~.
\label{sqrt}
\eeq
Although $D$ and $\bar{D}$ themselves are fermionic, their squares are 
thus bosonic operators
without any trace left of their fermionic origin. As we shall see shortly, 
this property is crucial in setting up a correct superfield formulation.

\noi
Let there now be given a Grassmann-odd BRST operator 
\mb{\hat{\Omega}=\hat{\Omega}(\hat{\zzzz}(t,\theta))}
and a BRST improved Hamiltonian 
\mb{\hat{H}=\hat{H}(\hat{\zzzz}(t,\theta))} with the properties that
\beq
[\hat{\Omega}(\hat{\zzzz}(t,\theta)),
\hat{\Omega}(\hat{\zzzz}(t,\theta))] ~=~ 0 
~~~~~~~~{\mbox{\rm and}}~~~~~~~~
[\hat{H}(\hat{\zzzz}(t,\theta)),
\hat{\Omega}(\hat{\zzzz}(t,\theta))] ~=~ 0~.
\label{homega}
\eeq
Without considering the superfield extension,
the existence of these two operators for an arbitrary admissible gauge
algebra (possibly open, of any rank, and possibly reducible, at any 
finite stage) associated with the first-class constraints is known from the
work of ref.\ \cite{BFV}. They are defined on the extended phase space
of fields that includes the required number of symplectic
ghost field pairs and Lagrange multipliers. Here we for the moment
assume that the required superfield extension is possible, and we shall
subsequently, in the next section, verify that this indeed is the case.

\noi
In our construction it turns out that a more fundamental r\^{o}le
is played by two Grassmann-odd operators, which are combinations of the
BRST operator and the Hamiltonian: 
\bea
\hat{Q}&=&\hat{Q}(\hat{\zzzz}(t,\theta),\theta)
~=~ \hat{\Omega}(\hat{\zzzz}(t,\theta))
 + \theta \hat{H}(\hat{\zzzz}(t,\theta)) ~, \cr
\hat{\bar{Q}}&=&\hat{\bar{Q}}(\hat{\zzzz}(t,\theta),\theta)
~=~ \hat{\Omega}(\hat{\zzzz}(t,\theta))
 - \theta \hat{H}(\hat{\zzzz}(t,\theta)) ~.
\label{qdef}
\eea
They are both nilpotent (anticommuting) by virtue of \eq{homega}:
\beq
[\hat{Q}, \hat{Q}] ~=~ 0 ~,~~~~~~~~~~
[\hat{Q}, \hat{\bar{Q}}] ~=~ 0 ~,~~~~~~~~~~
[\hat{\bar{Q}}, \hat{\bar{Q}}] ~=~ 0 ~,~~~~~~~~~~ ~.
\label{q}
\eeq
We demand that both $\hat{H}$ and $\hat{\Omega}$ be hermitian, a property
which is then inherited by $\hat{Q}$ and $\hat{\bar{\Omega}}$ if 
$\theta$ is real:
\beq
\hat{H}^{\dagger} ~=~ \hat{H} ~~~,~~~~~
\hat{\Omega}^{\dagger} ~=~ \hat{\Omega} ~~~,~~~~~
\hat{Q}^{\dagger} ~=~ \hat{Q}~~~,~~~~~
\hat{\bar{Q}}^{\dagger} ~=~ \hat{\bar{Q}} ~.\label{herm}
\eeq

\vspace{0.2cm}
\subsection{\sc Gauge Fixing}
\label{secgaugefix}

All of these relations hold before any gauge fixing. As is well known, 
for theories with first-class constraints, the supercommutation relations 
\e{homega} encode the gauge algebra of the theory, and we can view 
$\hat{H}$ and $\hat{\Omega}$ as generating operators of this gauge algebra.
The phase space operators $\hat{\zzzz}^A_0$ is of course extended to
include the full required set of ghost fields, and, in Darboux
coordinates, their canonically conjugate momenta. Their superfield
generalizations follow from \eq{gammasuper}.
To obtain well-defined dynamics, gauge fixing must be 
correctly implemented. According to the BFV theorem \cite{BFV},
this is accomplished with the help of
a gauge fermion $\hat{\Psi}_{1}$, whose supercommutator with $\hat{\Omega}$
is added to $\hat{H}$:
\beq
\hat{H}_{\hat{\Psi}_{1}} ~\equiv~ \hat{H} +
(i\hbar)^{-1}[\hat{\Psi}_{1},\hat{\Omega}] ~.\label{usualgaugefix}
\eeq
The subscript $1$ will be explained below.

\noi
Let us now consider the problem in the superfield formulation.
We propose to achieve the required gauge fixing by means of the 
following construction. Introduce two {\em bosonic} 
Hermitian operators $\hat{\Psi}'$ and  $\hat{\Psi}''$
(and the corresponding barred operators 
$\hat{\bar{\Psi}}'$ and $\hat{\bar{\Psi}}''$, where the
explicit $\theta$-dependence comes with the sign changed, 
similary to \eq{qdef}.): 
\bea
\hat{\Psi}'&=&\hat{\Psi}'(\hat{\zzzz}(t,\theta),\theta)
~=~ \hat{\Psi}'_{0}(\hat{\zzzz}(t,\theta)) 
+ \theta\hat{\Psi}'_{1}(\hat{\zzzz}(t,\theta))~, \cr
\hat{\Psi}''&=&\hat{\Psi}''(\hat{\zzzz}(t,\theta),\theta)
~=~ \hat{\Psi}''_{0}(\hat{\zzzz}(t,\theta)) 
+ \theta\hat{\Psi}''_{1}(\hat{\zzzz}(t,\theta)) ~. 
\eea
They are composed of, in total, four components, 
$\hat{\Psi}'_{0}$, $\hat{\Psi}'_{1}$, $\hat{\Psi}''_{0}$ 
and $\hat{\Psi}''_{1}$, of the following statistics:
\beq
\epsilon(\hat{\Psi}'_{0}) ~=~ 0~=~\epsilon(\hat{\Psi}''_{0}) ~,~~~~~~~~~
\epsilon(\hat{\Psi}'_{1})~=~1~=~\epsilon(\hat{\Psi}''_{1})~.
\eeq
We shall see below that a combination of the one-components 
$\hat{\Psi}'_{1}$ and $\hat{\Psi}''_{1}$ comprise the usual gauge 
fermion of the BFV construction when \mb{\hat{\Psi}_{0}=0}.
This explains why we put the subindex $1$ in \eq{usualgaugefix}.

\noi
For notational convenience, let us denote $\hat{\Psi}'_{0}$ by just
$\hat{\Psi}_0$. The relevant nilpotent BRST operator is
\beq
\hat{\Omega}_{\hat{\Psi}_{0}}
 ~=~ e^{(i\hbar)^{-1}{\rm ad}\hat{\Psi}_{0}}\hat{\Omega}
 ~=~ e^{(i\hbar)^{-1}\hat{\Psi}_{0}}~\hat{\Omega}~
e^{-(i\hbar)^{-1}\hat{\Psi}_{0}} ~,\label{opsio}
\eeq
where we have let
\beq
{\rm ad}  \hat{X} ~\equiv~ [\hat{X},~\cdot~]
\eeq
denote the adjoint action of an operator $\hat{X}$. Next, choose
a $\hat{\Psi}$-dressed $Q$-operator be a sum of an ``exponential" 
and a ``linear" gauge-fixing part:
\bea
\hat{Q}_{\hat{\Psi}}
&=&\hat{Q}_{\hat{\Psi}}(\hat{\zzzz}(t,\theta),\theta) \cr
 &=& e^{(i\hbar)^{-1} {\rm ad} \hat{\Psi}'(\hat{\zzzz}(t,\theta),\theta)}
\hat{Q}(\hat{\zzzz}(t,\theta),\theta) \cr
&& +(i\hbar)^{-1} [ \hat{\Psi}''(\hat{\zzzz}(t,\theta),\theta) ,
\hat{\Omega}_{\hat{\Psi}_{0}}(\hat{\zzzz}(t,\theta)) ] 
~.\label{qdef1}
\eea
We shall later, in Section \ref{seceqofmot}, see that 
$\hat{Q}_{\hat{\Psi}}$ plays the r\^{o}le of a
``super-Hamiltonian'' \ie an evolution operator for the superfields.
It can be written in terms of two components:
\beq
  \hat{Q}_{\hat{\Psi}}(\hat{\zzzz}(t,\theta),\theta)~=~
\hat{\Omega}_{\hat{\Psi}}(\hat{\zzzz}(t,\theta))
+ \theta \hat{H}_{\hat{\Psi}}(\hat{\zzzz}(t,\theta))~,
\eeq
where
\bea
\hat{\Omega}_{\hat{\Psi}}
&=&\hat{\Omega}_{\hat{\Psi}}(\hat{\zzzz}(t,\theta))
~\equiv~\hat{Q}_{\hat{\Psi}}(\hat{\zzzz}(t,\theta),0)~, \cr
\hat{H}_{\hat{\Psi}}
&=&\hat{H}_{\hat{\Psi}}(\hat{\zzzz}(t,\theta))
~\equiv~{{\partial} \over {\partial\theta}} 
\hat{Q}_{\hat{\Psi}}(\hat{\zzzz}(t,\theta),\theta)~.
\eea
Inserting the definition (\ref{qdef}) into \eq{qdef1}, we see that
going from the operator $\hat{Q}$ to
$\hat{Q}_{\hat{\Psi}}$ corresponds to the
following modifications of  $\hat{\Omega}$ and $\hat{H}$:
\beq
\hat{\Omega}_{\hat{\Psi}} ~=~ \hat{\Omega}_{\hat{\Psi}_{0}}
 +(i\hbar)^{-1} [ \hat{\Psi}''_{0} ,
\hat{\Omega}_{\hat{\Psi}_{0}} ]  ~,\label{opsi}
\eeq
and
\bea
\hat{H}_{\hat{\Psi}} &=&  
e^{(i\hbar)^{-1}{\rm ad}\hat{\Psi}_{0}}\left( \hat{H}
 + (i\hbar)^{-1} [\int_{0}^{1}\! d\alpha\left(
e^{-\alpha(i\hbar)^{-1}{\rm ad}\hat{\Psi}_{0}}\hat{\Psi}'_{1}\right)
+ e^{-(i\hbar)^{-1}{\rm ad}\hat{\Psi}_{0}} \hat{\Psi}''_{1} ,
\hat{\Omega} ]   \right) \cr
&=& e^{(i\hbar)^{-1}{\rm ad}\hat{\Psi}_{0}}\hat{H}
 + (i\hbar)^{-1}[\int_{0}^{1}\! d\alpha\left(
e^{+\alpha(i\hbar)^{-1}{\rm ad}\hat{\Psi}_{0}}\hat{\Psi}'_{1}\right) +
\hat{\Psi}''_{1} ,
\hat{\Omega}_{\hat{\Psi}_{0}} ]  \cr
&=&  e^{(i\hbar)^{-1}{\rm ad}\hat{\Psi}_{0}}\hat{H} 
+ (i\hbar)^{-1}[\hEo(\hat{\Psi}'_{1})
+ \hat{\Psi}''_{1} ,\hat{\Omega}_{\hat{\Psi}_{0}}]  ~.
\label{hpsi}
\eea
We have here used the shorthand notation of
\beq
\hEo( \hat{X} ) ~\equiv~ \int_{0}^{1}\! d\alpha~ 
e^{+\alpha (i\hbar)^{-1} {\rm ad}\hat{\Psi}_{0}}\hat{X}
 ~=~f\left( (i\hbar)^{-1} {\rm ad}\hat{\Psi}_{0} \right)\hat{X}~,\label{edef}
\eeq
where
\beq
 f(x)~=~ \int_{0}^{1} d\alpha \
e^{\alpha x}~=~\frac{e^{x}-1}{x}
~=~\sum_{n=1}^{\infty}\frac{x^{n-1}}{n!}~.
\eeq
It follows from \eqs{opsi}{hpsi} that the definition
(\ref{qdef1}) has the following effects: 
\begin{itemize}
\item
\mb{\hat{\Psi}_{0}\equiv \hat{\Psi}'_{0}} induces a canonical 
transformation on both $\hat{\Omega}$ and $\hat{H}$. 
\item
Both $\hat{\Psi}'_{1}$ and $\hat{\Psi}''_{1}$ induce 
gauge-fixing terms in the Hamiltonian. {}From now on we will therefore call  
$\hat{H}_{\Psi}$ of \eq{hpsi} the superfield {\em unitarizing Hamiltonian}.
\item
$\hat{\Psi}''_{0}$ adds a term to the zero-component 
$\hat{\Omega}_{\hat{\Psi}}$, so that it in general  is different
from the pertinent BRST operator $\hat{\Omega}_{\hat{\Psi}_{0}}$.
\end{itemize}

\noi
All hermiticity properties \e{herm} are inherited by the 
$\hat{\Psi}$-dressed operators. This follows from the fact that the superfield
$\hat{\Psi}$ must be hermitian (consistent with the fact that the 
original ``gauge fermion'' $\hat{\Psi}_1$ must be chosen antihermitian):
\beq
\hat{\Psi}^{\dagger} ~=~ \hat{\Psi} ~~,~~~~~~ \hat{\Psi}_1^{\dagger}
~=~ -\hat{\Psi}_1 ~.
\eeq

\noi
The operator $\hEo$ is invertible:
\beq
\hEo^{-1}( \hat{X} )~=~ \left(\frac{1}{f}\right)\left( 
(i\hbar)^{-1} {\rm ad}\hat{\Psi}_{0} \right)\hat{X}~,
\eeq
where
\beq
\frac{1}{f(x)} ~=~  \frac{x}{e^{x}-1}
~=~ \sum_{n=0}^{\infty} B_{n} \frac{x^n}{n!}
 ~=~ 1 - \frac{x}{2} -
\sum_{k=1}^{\infty} B'_{k}  \frac{(-x^{2})^{k}}{(2k)!} ~,
\eeq
is the generating function for the Bernoulli numbers.\footnote{
Note that there are two different conventions for the 
Bernoulli numbers in the literature.}
This shows that one can change the $1$-component of the gauge-fixing
from a ``exponential'' gauge to a ``linear'' gauge 
and vice-versa. This is an important property, and it shows that one of 
the four gauge-fixing components mentioned above 
in fact is redundant. However, depending on the application, different forms 
of the gauge-fixing can prove convenient. 

\noi
The ``super-Hamiltonian'' $\hat{Q}_{\hat{\Psi}}$ is BRST invariant:
\beq
[\hat{Q}_{\hat{\Psi}}(\hat{\zzzz}(t,\theta),\theta),
\hat{\Omega}_{\hat{\Psi}_{0}}(\hat{\zzzz}(t,\theta)] ~=~ 0
\label{superhambrsinv} ~.
\eeq 
This leads to the fact that the BRST operator 
\mb{\hat{\Omega}_{\hat{\Psi}_{0}}(\hat{\zzzz}(t,\theta))} is a constant 
of motion; see also the discussion in Section 
\ref{seceqofmot}-\ref{seconofmot}.

\noi
For consistency (the equations of motion should be integrable, 
see Section \ref{seceqofmot}) 
we require that the ``super-Hamiltonian'' is nilpotent:
\beq
[\hat{Q}_{\hat{\Psi}}(\hat{\zzzz}(t,\theta),\theta),
\hat{Q}_{\hat{\Psi}}(\hat{\zzzz}(t,\theta),\theta)] 
~=~ 0 ~.\label{qnilp}
\eeq
Remarkably, this crucial condition yields both of the equations
\beq
[\hat{\Omega}_{\hat{\Psi}}(\hat{\zzzz}(t,\theta)),
\hat{\Omega}_{\hat{\Psi}}(\hat{\zzzz}(t,\theta))] ~=~ 0 
~~~~~~~~{\mbox{\rm and}}~~~~~~~~~
[\hat{H}_{\hat{\Psi}}(\hat{\zzzz}(t,\theta)),
\hat{\Omega}_{\hat{\Psi}}(\hat{\zzzz}(t,\theta))] ~=~ 0 ~.
\label{homega1}
\eeq
Conversely, they of course imply the nilpotency condition \e{qnilp}. 
The conditions (\ref{homega1}) are automatically satisfied for gauges with 
\mb{\hat{\Psi}''_{0} = 0}. Although there are non-trivial gauges
with \mb{\hat{\Psi}''_{0} \neq  0} fulfilling this consistency requirement
\e{homega1}, we shall for simplicity from now on limit ourselves
to the situation where \mb{\hat{\Psi}''_{0} = 0}. 
Then $\hat{\Omega}_{\hat{\Psi}}$ coincides with the BRST operator
$\hat{\Omega}_{\hat{\Psi}_{0}}$, and the unitarizing operators
can, as explained earlier, be written entirely with the help of 
the ``exponential'' form of gauge fixing. However, most of what follows
can easily be extended to include consistent \mb{\hat{\Psi}''_{0} 
\neq  0} gauges.

\vspace{0.2cm}
\subsection{\sc Equations of Motion}
\label{seceqofmot}

Consider now the following superfield equation of motion:
\beq
i\hbar D\hat{\zzzz}^{A}(t,\theta) 
~=~ -[\hat{Q}_{\hat{\Psi}}(\hat{\zzzz}(t,\theta),\theta),
\hat{\zzzz}^{A}(t,\theta)] ~.
\label{sh}
\eeq
By applying the $D$-operator again, and
taking into account the explicit $\theta$-dependence on the r.h.s.\ of
\eq{sh} when performing the second differentiation,
as well as using the Jacobi identity for the supercommutator, one finds,
remarkably,
\beq
i\hbar \frac{d}{dt}\hat{\zzzz}^{A}(t,\theta) ~=~ -[\hat{H}_{\hat{\Psi}}
(\hat{\zzzz}(t,\theta)),\hat{\zzzz}^{A}(t,\theta)] ~.
\label{sh1}
\eeq
In fact, this is just the anticipated form of the Heisenberg equation of
motion for the superfield $\hat{\zzzz}^{A}$ and the unitarizing Hamiltonian
$\hat{H}_{\hat{\Psi}}$. It is an exact superfield operator relation. 

\noi
Multiplying both sides of \eq{sh1} by $\theta$ from the left,
and substituting into \eq{sh}, one finds that the other independent
relation contained in (\ref{sh}) is
\beq
i\hbar\frac{d}{d\theta}\hat{\zzzz}^{A}(t,\theta) ~=~ 
-[\hat{\Omega}_{\hat{\Psi}}(\hat{\zzzz}(t,\theta)),
\hat{\zzzz}^A(t,\theta)] ~. 
\label{shomega}
\eeq
Thus, while the ordinary time evolution is generated by the Hamiltonian
$\hat{H}_{\hat{\Psi}}$, the evolution in the Grassmann-odd direction 
is generated by the operator $\hat{\Omega}_{\hat{\Psi}}$. 
The condition 
\mb{(d/d\theta)^2\hat{\zzzz}^{A}(t,\theta) = 0} is consistently 
reproduced from the right hand side of \eq{shomega} on account of the 
nilpotency condition on $\hat{\Omega}_{\hat{\Psi}}$ from \eq{homega}. 
Conversely, {\em in a superspace formulation,
the nilpotency condition on the BRST charge can be viewed as a trivial
consequence of the fermionic nature of the $\theta$-direction.} 
(Remarkably, due solely to the presence of the explicit 
$\theta$-dependence in the definition (\ref{qdef}), the nilpotency
condition on $\hat{Q}_{\hat{\Psi}}$, \eq{q}, does not lead to 
the analogous and would-be fatal conclusion that the right hand side 
of \eq{sh1} should vanish.)

\noi
By differentiating \eq{sh1} with respect to $\theta$, and
differentiating \eq{shomega} with respect to $t$, we see that the
integrability condition 
\mb{[d/dt, d/d\theta] \hat{\zzzz}^{A}(t,\theta) = 0} 
is satisfied on account of the second equation in \e{homega1},
and because of the Jacobi identity for supercommutators.

\vspace{0.2cm}
\subsection{\sc Constants of Motion}
\label{seconofmot}

Let us now analyze the two eqs.\ of motion (\ref{sh1}) and (\ref{shomega})
in more detail. First, although \eq{sh1} looks deceptively like
the correct quantum mechanical equation for the original phase space
operators $\hat{\zzzz}_{0}^{A}(t)$, it should be recalled that both 
components of the superfields $\hat{\zzzz}^A(t,\theta)$ are involved,
and that they in principle could mix through 
$\hat{H}_{\hat{\Psi}}(\zzzz(t,\theta))$.
Consider, however, the $\theta$-dependence of the unitarizing Hamiltonian.
We find it by taking the derivative, and using \eq{shomega}:
\bea
\frac{d}{d\theta}\hat{H}_{\hat{\Psi}}(\hat{\zzzz}(t,\theta))  
~=~ -(i\hbar)^{-1}[\hat{\Omega}_{\hat{\Psi}}(\hat{\zzzz}(t,\theta)),
\hat{H}_{\hat{\Psi}}(\hat{\zzzz}(t,\theta))] ~=~ 0 ~, \label{hthindep}
\eea
where the last equality follows from \eq{homega}. Thus, the operator
$\hat{H}_{\hat{\Psi}}(\zzzz(t,\theta))$ is, despite appearances, 
$\theta$-independent. Similarly, one finds that
\beq
\frac{d}{d\theta} \hat{\Omega}_{\hat{\Psi}}(\hat{\zzzz}(t,\theta))
 ~=~ -(i\hbar)^{-1}[\hat{\Omega}_{\hat{\Psi}}(\hat{\zzzz}(t,\theta)),
 \hat{\Omega}_{\hat{\Psi}}(\hat{\zzzz}(t,\theta))] ~=~ 0 ~.\label{othindep}
\eeq
Also the operator $\hat{\Omega}_{\hat{\Psi}}(\zzzz(t,\theta))$ is therefore 
$\theta$-independent.
The corresponding argument applied to \eq{sh1} shows that, as
expected, both the unitarizing Hamiltonian $\hat{H}_{\hat{\Psi}}$ and the 
operator $\hat{\Omega}_{\hat{\Psi}}$ are constants of motion:
\bea
\frac{d}{dt}\hat{H}_{\hat{\Psi}}(\hat{\zzzz}(t,\theta)) 
&~=~& - (i\hbar)^{-1}[\hat{H}_{\hat{\Psi}}(\hat{\zzzz}(t,\theta)),
\hat{H}_{\hat{\Psi}}(\hat{\zzzz}(t,\theta))] ~=~ 0 \cr
\frac{d}{dt}\hat{\Omega}_{\hat{\Psi}}(\hat{\zzzz}(t,\theta)) 
&~=~& -(i\hbar)^{-1}[\hat{H}_{\hat{\Psi}}(\hat{\zzzz}(t,\theta)),
\hat{\Omega}_{\hat{\Psi}}(\hat{\zzzz}(t,\theta))] ~=~ 0 ~.
\eea

\noi
One important consequence of the above results is the following. Since both
$\hat{\Omega}_{\Psi}$ and $\hat{H}_{\Psi}$ are $\theta$-independent, and
hence in fact only functions of the zero-components $\hat{\zzzz}^{A}_{0}$,
the fundamental supercommutator relations between them can be satisfied
by the conventional solutions \cite{BFV}, once $\hat{\Psi}$-dressed in
the manner described above. The existence of appropriate solutions to
the fundamental supercommutation relation (\ref{qnilp}) is therefore 
guaranteed. 

\vspace{0.2cm}
\subsection{\sc Component Formulation}

\noi
Letting \mb{\theta=0}, the two superfield equations of motion 
\e{sh1} and \e{shomega} reduce to
\beq
i\hbar\frac{d}{dt} \hat{\zzzz}_{0}^{A}(t) 
~=~ -[\hat{H}_{\hat{\Psi}} (\hat{\zzzz}_{0}(t)),
\hat{\zzzz}_{0}^{A}(t)]
\label{eqofmotham0}
\eeq
and
\beq
i\hbar\hat{\zzzz}_{1}^{A}(t)
 ~=~ -[\hat{\Omega}_{\hat{\Psi}}(\hat{\zzzz}_{0}(t)),
\hat{\zzzz}_{0}^{A}(t)] ~,
\label{eqofmotomega0}
\eeq
respectively. The former \e{eqofmotham0} of these relations is precisely the 
correct Heisenberg eq.\ of motion for the original phase space 
operators $\hat{\zzzz}_{0}^{A}(t)$ with respect to the 
unitarizing Hamiltonian $\hat{H}_{\hat{\Psi}}$. 
The latter relation \e{eqofmotomega0} shows that the corresponding 
superfield partners $\hat{\zzzz}_{1}^{A}(t)$ are given, 
{\em through an exact operator identity}, 
by the $\hat{\Omega}_{\hat{\Psi}}$-transform of the original 
$\hat{\zzzz}_{0}^{A}$'s. 

\noi 
The two equations \e{eqofmotham0} and \e{eqofmotomega0} 
are in fact equivalent to \e{sh1} and \e{shomega}.
This is trivial for the former equation \e{sh1}
by performing a super translation of \e{eqofmotham0}.
The latter equation follows from two observations:
First, by \e{eqofmotomega0}, the Jacobi identity and nilpotency
of \mb{\hat{\Omega}_{\hat{\Psi}}(\hat{\zzzz}_{0}(t))}
\beq
  [\hat{\zzzz}_{1}^{A}(t),
\hat{\Omega}_{\hat{\Psi}}(\hat{\zzzz}_{0}(t))]~=~0~.
\eeq
Second, \mb{\hat{\Omega}_{\hat{\Psi}}(\hat{\zzzz}(t,\theta))} 
is independent of $\theta$:
\beq
 \hat{\Omega}_{\hat{\Psi}}(\hat{\zzzz}(t,\theta))
~=~\hat{\Omega}_{\hat{\Psi}}\left( 
\hat{U}^{-1}(t)~\hat{\zzzz}_{0}(t)~ \hat{U}(t)\right)
~=~\hat{U}^{-1}(t)~\hat{\Omega}_{\hat{\Psi}}
(\hat{\zzzz}_{0}(t))~ \hat{U}(t)
~=~ \hat{\Omega}_{\hat{\Psi}}(\hat{\zzzz}_{0}(t))~,
\eeq 
for 
\beq
\hat{U}(t)\equiv \exp\left[(i\hbar)^{-1}\theta~ 
\hat{\Omega}_{\hat{\Psi}}(\hat{\zzzz}_{0}(t)) \right] ~.
\eeq
Now eq. \e{shomega} follows quite easily.

\vspace{0.2cm}
\subsection{\sc Ghost Number Assignments}
\label{secghostnumber}

At this point it is appropriate to briefly comment on the question
of ghost number assignments. As is well known, the operator 
$\hat{\Omega}$ (and hence also $\hat{\Omega}_{\hat{\Psi}_{0}}$) 
is not only Grassmann-odd, but also carries ghost number of one
unit: \mb{{\rm gh}(\hat{\Omega}) = +1}. 
Treating \eq{shomega} as a first principle, 
we define \mb{{\rm gh}(\theta) = -1}.
Such an assignment is totally compatible with 
the other eq.\ of motion, (\ref{sh1}). 
However now the operators $D$, $\hat{\Psi}$, $\hat{Q}$ and 
$\hat{Q}_{\hat{\Psi}}$ carry no definite ghost number.
In fact, the ghost number discrepancy between the two supercomponent
parts is the same for all these operators, namely two units, cf.\
\beq
 {\rm gh}(\hat{\Omega}) = +1~,~~~~~~
{\rm gh}(\hat{H}) = 0~,~~~~~~
{\rm gh}(\hat{\Psi}_{0}) = 0~,~~~~~~
{\rm gh}(\hat{\Psi}_{1}) = -1~.
\eeq
However, considering the way in
which the defining equation (\ref{sh}) splits up into the two independent
equations (\ref{sh1}) and (\ref{shomega}), it is immediately clear that
this indefinite ghost number of $D$ and, say, $\hat{Q}_{\hat{\Psi}}$, 
is of absolutely no consequence. The mismatch can be easily corrected,
for example,
by inserting a bosonic constant $\eta$ carring ghost number +2 in front of the 
one-component of the above mentioned operators:
\beq
\begin{array}{rcccl}
D &=& \frac{d}{d\theta} &+& \eta \theta\frac{d}{dt} ~,\\
\hat{\Psi} &=& \hat{\Psi}_{0} &+&  \eta \theta \hat{\Psi}_{1} ~, \\
\hat{Q} &=& \hat{\Omega} &+&  \eta \theta \hat{H} ~, \\
\hat{Q}_{\hat{\Psi}} &=& \hat{\Omega}_{\hat{\Psi}} 
&+&  \eta \theta \hat{H}_{\hat{\Psi}}~.
\end{array}
\eeq
Note that $D$
is still essentially the square root of the time derivative:
\beq
D^2 ~=~ \eta \frac{d}{dt} ~.
\eeq

\noi
Let us finally mention that, for consistency, one must add to the usual
ghost number operator \cite{BFV} a term $+2\eta\partial_{\eta}$. Since all
these redefinitions are inessential, we prefer, however, to avoid them.

\vspace{0.2cm}
\subsection{\sc The Physical-State Condition and Gauge Independence}

To define physical states, we require that 
\mb{|{\rm phys}\rangle_{\hat{\Psi}_{0}}}
be annihilated by $\hat{\Omega}_{\hat{\Psi}_{0}}(t,\theta)$:
\beq
\hat{\Omega}_{\hat{\Psi}_{0}}
|{\rm phys}\rangle_{\hat{\Psi}_{0}} 
~=~ 0 ~. \label{phys}
\eeq
By hermiticity of $\hat{\Omega}_{\hat{\Psi}_{0}}(t,\theta)$, 
this is equivalent to
\beq
 {}_{\hat{\Psi}_{0}}\! \langle{\rm phys}| 
\hat{\Omega}_{\hat{\Psi}_{0}}
~=~ 0 ~.
\eeq
These two conditions are identical to the usual requirements for physical
states \cite{KO}. Note that we have labelled our physical states by the
subscript $\hat{\Psi}_{0}$. This is because, as we have seen above, this part
of the superfield gauge boson $\hat{\Psi}$ is responsible for a canonical
transformation, and the corresponding physical states must be rotated
accordingly. Conventionally one chooses 
\mb{\hat{\Psi}_{0} = 0} from the outset,
but the above definition is the suitable one in our more general case.

\noi
For the reader who is puzzled by the fact that only
the original $\hat{\Omega}_{\hat{\Psi}_{0}}$ 
BRST operator enters in the definition of
physical states, we note that in the case of exponential gauge fixing
\mb{\Psi=\Psi'=\Psi'_{0}+\theta\Psi'_{1}} with \mb{\Psi''=0},
it trivial to rewrite the physical-state condition in terms of
$\hat{Q}_{\hat{\Psi}}$ and $\hat{\bar{Q}}_{\hat{\Psi}}$, 
which one would have guessed 
should enter into the definition of physical BRST cohomology. Indeed,
\beq
\left(\hat{Q}_{\hat{\Psi}} + 
\hat{\bar{Q}}_{\hat{\Psi}}\right)
|{\rm phys}\rangle_{\hat{\Psi}_{0}} ~=~ 0 ~.
\eeq

\noi
Corresponding to the definition of physical states, we also define 
physical operators 
$\hat{A}_{(\hat{\Psi})}(\hat{\zzzz}(t,\theta),\theta)$ 
as those that supercommute with
$\hat{\Omega}_{\hat{\Psi}_{0}}(\hat{\zzzz}(t,\theta))$:
\beq
[\hat{A}_{(\hat{\Psi})}(\hat{\zzzz}(t,\theta),\theta),
\hat{\Omega}_{\hat{\Psi}_{0}}(\hat{\zzzz}(t,\theta))] ~=~ 0 ~.
 \label{physo}
\eeq
We have here explicitly given the operator 
\mb{\hat{A}_{(\hat{\Psi})}(\hat{\zzzz}(t,\theta),\theta)} 
a subscript $\hat{\Psi}$ in order to emphasize that the operator
\beq
\hat{A}_{(\hat{\Psi})}(\hat{\zzzz}(t,\theta),\theta)
~=~\hat{U}_{\hat{\Psi}}(t,\theta)~
\hat{A}(\hat{\zzzz}(t_{i},\theta_{i}),\theta_{i})
~\hat{U}_{\hat{\Psi}}^{-1} (t,\theta)
\eeq
is evolving from initial conditions at \mb{t_{i}} and \mb{\theta_{i}=0}
according to Heisenberg equation of motion by an evolution operator
\beq
 \hat{U}_{\hat{\Psi}}(t,\theta) ~=~T \exp\left[ I\left( 
 -(i\hbar)^{-1}\hat{Q}_{\hat{\Psi}}+ {\cal D} \right)
  \right](t,\theta)~
\eeq
depending on the gauge $\hat{\Psi}$. Here $T$ stands for time-ordering
with respect to time $t$ and we have introduced the operator \mb{I=D^{-1}}, 
inverse to $D$. It can be written as an integral operator
\beq
I[F](t,\theta) ~\equiv~ 
\int \! dt' d\theta' ~K(t,\theta;t',\theta')~F(t',\theta')
\eeq
with the kernel
\beq
K(t,\theta;t',\theta') ~\equiv~ 
1_{[t_{i},t]}(t')~-~\theta ~\delta(t-t')~ \delta(\theta') ~.
\eeq
Here \mb{1_{[t_{i},t]}(t')} is just the characteristic function, \ie,
\beq
 1_{[t_{i},t]}(t')~=~1~~ {\rm if}~~~ t_{i} \leq t'  \leq t~,
\eeq
and zero otherwise.

\noi
The above physicality definitions (\ref{phys}) and (\ref{physo})
can now be used to provide a proof of gauge independence of physical 
matrix elements 
$$
 {}_{\hat{\Psi}_{0}}\! \langle{\rm phys}| 
 \hat{A}_{(\hat{\Psi})}(\hat{\zzzz}(t,\theta),\theta) 
 |{\rm phys}'\rangle_{\hat{\Psi}_{0}} 
$$
in superfield language. First, since
$\hat{\Psi}_{0}$ simply generates canonical transformations, 
the picture changes with the choice of $\hat{\Psi}_{0}$, 
and it does not make sense to try to change $\hat{\Psi}_{0}$ 
in a given fixed $\hat{\Psi}_{0}$-picture.
Indeed, the whole formalism is invariant under canonical 
transformations, provided we redefine the physical states
accordingly, as explained above.  
Without any loss of generality we can therefore restrict
ourselves to changes 
\mb{\hat{\Psi} \to \hat{\Psi}+\Delta\hat{\Psi}} with
\beq
\Delta\hat{\Psi} ~=~ \theta\Delta\hat{\Psi}_{1}\label{deltapsi} 
\eeq
keeping zero component $\hat{\Psi}_{0}$ fixed.
This induces a change in $\hat{Q}_{\hat{\Psi}}$ of the 
form 
\beq
\Delta\hat{Q}_{\hat{\Psi}} ~=~ (i\hbar)^{-1}
 [ \hEo(\Delta\hat{\Psi}),\hat{\Omega}_{\hat{\Psi}_{0}}] ~.
\eeq
The change in $\hat{Q}_{\hat{\Psi}}$ is thus effectively that of a BRST
supercommutator. This is the crucial property, shared with the conventional
treatment \cite{BFV}, which allows us to demonstrate independence of the
chosen gauge.

\noi
We now introduce a ``gauge changing'' operator \cite{BF}
\beq
\hat{G}(t,\theta)~=~  \hat{U}_{\hat{\Psi}+\Delta\hat{\Psi}}(t,\theta)
 ~\hat{U}_{\hat{\Psi}}^{-1} (t,\theta)
~=~ T \exp\left[-(i\hbar)^{-1}I~\Delta\hat{Q}_{\hat{\Psi}}
 \right] (t,\theta) ~.
\eeq  
It satisfies the following equation of motion,
\beq
i\hbar D\hat{G}(t,\theta) ~=~
 - \Delta\hat{Q}_{\hat{\Psi}}(t,\theta)~\hat{G}(t,\theta) ~,
\label{gequation}
\eeq
subject to the boundary condition
\beq
\hat{G}(t\!=\!t_{i},\theta\!=\! \theta_{i}) ~=~ \hat{\bf 1} ~.
\eeq

\noi
Consider now the solutions of the equation of motion (\ref{sh}) with
respect to the new $\hat{Q}$-operator $\hat{Q}_{\hat{\Psi}+\Delta\hat{\Psi}}$.
They are related to the previous solutions through
\beq
\hat{\zzzz}^A_{(\hat{\Psi}+\Delta\hat{\Psi})} ~=~ \hat{G}~\hat{
\zzzz}^A_{(\hat{\Psi})}~\hat{G}^{-1} ~.
\eeq
In general, for arbitrary operators $\hat{A}$,
\beq
\hat{A}_{(\hat{\Psi}+\Delta\hat{\Psi})} ~=~ \hat{G}~\hat{A}_{(\hat{\Psi})}
~\hat{G}^{-1} ~. \label{opchange}
\eeq

\noi
Let us now restrict ourselves to an infinitesimal change of superfield 
gauge bosons, for which the above solution for $\hat{G}$ simply reads
\beq
\hat{G} ~=~ \hat{\bf 1}+\delta\hat{G}~,~~~~~~~~~~
\delta\hat{G}~=~ - (i\hbar)^{-2}\!I\left[\hEo(\delta
\hat{\Psi}),\hat{\Omega}_{\hat{\Psi}_{0}} \right] 
 ~=~  - (i\hbar)^{-2}\!\left[I[\hEo(\delta
\hat{\Psi})],\hat{\Omega}_{\hat{\Psi}_{0}}\right] ~,
\eeq
due to the fact that $\hat{\Omega}_{\hat{\Psi}_{0}}$ 
is a constant of motion in both the $t$ and $\theta$ directions.
This fact also implies  
that the BRST charge $\hat{\Omega}_{\hat{\Psi}_{0}}(t,\theta)$ 
away from the initial Cauchy surface at initial time
 \mb{t_{i}} and \mb{\theta_{i}=0}
{\em does not depend on the gauge} $\hat{\Psi}$.
The relation (\ref{opchange}) becomes
\beq
\delta \hat{A}_{(\hat{\Psi})} ~=~ [\delta\hat{G},\hat{A}_{(\hat{\Psi})}] ~.
\eeq
As a nice check, we note explicitly, that, as it should,
\beq
\delta\left(\hat{\Omega}_{\hat{\Psi}_{0}}\right)_{(\hat{\Psi})}
 ~=~ [\delta\hat{G}, \hat{\Omega}_{\hat{\Psi}_{0}}] ~=~ 0 ~.
\eeq

\noi
For an arbitrary operator $\hat{A}$ we get, using equation
(\ref{physo}),
\bea
\delta \hat{A}_{(\hat{\Psi})} 
&=& - [\hat{A}_{(\hat{\Psi})},\delta\hat{G}] \cr
&=& (i\hbar)^{-2}\!\left[\hat{A}_{(\hat{\Psi})},[I~
\hEo(\delta\hat{\Psi}),\hat{\Omega}_{\hat{\Psi}_{0}}]\right] \cr
&=& (i\hbar)^{-2}\!\left[[\hat{A}_{(\hat{\Psi})},I~
\hEo(\delta\hat{\Psi})],\hat{\Omega}_{\hat{\Psi}_{0}}\right] ~,
\eea
which is BRST-exact. Thus,
\beq
 {}_{\hat{\Psi}_{0}}\! \langle{\rm phys}| 
 \delta\hat{A}_{(\hat{\Psi})}(\hat{\zzzz}(t,\theta),\theta) 
 |{\rm phys}'\rangle_{\hat{\Psi}_{0}} ~=~ 0 ~.
\eeq
This shows that physical matrix elements do not depend on the chosen
gauge. 
Following the principles laid out in refs.\ \cite{KO} and \cite{BF0}, one
can analyze the condition of physical unitarity along similar lines.

\vspace{0.2cm}
\subsection{\sc Ward Identities}

The BRST symmetry gives rise to a superfield 
formulation of Ward identities derivable from this operator formulation. 
To see this, we proceed as in the usual
case \cite{BF} by introducing for the superfield operators
\mb{\hat{\zzzz}^A} external c-number sources, now themselves also
superfields: \mb{J_{A}(t,\theta) = J^{0}_{A}(t) + \theta J^{1}_{A}(t)}, 
of opposite statistics: \mb{\epsilon(J_{A}) = \epsilon_{A}+1}. 
We take the opportunity here to be slightly more
general, and introduce as well c-number sources, again superfields,
\mb{\zzzz^{*}_{A}(t,\theta) = \zzzz^{0*}_{A}(t) + \theta\zzzz^{1*}_{A}(t)}, 
with the {\em same statistics}: \mb{\epsilon(\zzzz^{*}_{A}) = \epsilon_{A}}. 
These additional superfields are 
to be sources of BRST-transforms of the superfield
phase-space variables \cite{BF0}.
Let us first introduce, for solutions $\hat{\zzzz}^A$ of \eq{sh}, 
a generating operator 
\mb{\hat{\cal Z}(t,\theta)=\hat{\cal Z}(J,\zzzz^{*};t,\theta)} 
satisfying the following equation:
\beq
i\hbar D\hat{\cal Z}(t,\theta)
~=~  \Delta\hat{Q}(t,\theta)~ \hat{\cal Z}(t,\theta) ~,\label{zeq}
\eeq
where
\beq
 - \Delta\hat{Q}(t,\theta) ~~\equiv~~ (-1)^{\epsilon_{A}+1}
 J_{A}(t,\theta)~\hat{\zzzz}^{A}(t,\theta) 
+(-1)^{\epsilon_{A}} (i\hbar)^{-1}\zzzz^{*}_{A}(t,\theta)~
[\hat{\zzzz}^{A}(t,\theta),\hat{\Omega}_{\hat{\Psi}_{0}}
(\hat{\zzzz}^{A}(t,\theta))]~, 
\eeq
and subject to the initial boundary condition
at time \mb{t_{i}} and \mb{\theta_{i}=0} 
\beq
\hat{\cal Z}(t \! = \! t_{i},
\theta \! = \! \theta_{i}) ~=~ \hat{\bf 1} ~.\label{zbc}
\eeq
We can write the solution as
\beq
\hat{\cal Z}(t, \theta)
~=~ T \exp\left[  (i\hbar)^{-1} I~ \Delta\hat{Q} \right] (t,\theta)
~.\label{zsol}
\eeq
Note that there is no explicit $t$ or $\theta$ dependence inside 
$\hat{\cal Z}(t, \theta)$.

\noi
The corresponding operator that interpolates between two events 
\mb{ t_{1}, \theta_{1}}  and \mb{t_{2}, \theta_{2}} is given as
\beq
\hat{\cal Z}_{21} 
~\equiv~\hat{\cal Z}(t_{2}, \theta_{2}; t_{1}, \theta_{1})
 ~=~ \hat{\cal Z}(t_{2}, \theta_{2})
~\hat{\cal Z}^{-1}(t_{1}, \theta_{1}) ~.
\eeq
It satisfies 
\beq
 \hat{\cal Z}_{11}=\hat{\bf 1}~,~~~~~~~ 
\hat{\cal Z}_{12}= \hat{\cal Z}_{21}^{-1} 
 ~,~~~~~~~  \hat{\cal Z}_{32} \hat{\cal Z}_{21} = \hat{\cal Z}_{31}~.
\eeq
We shall especially be interested in the full evolution
\beq
  \hat{\cal Z}_{fi} ~=~ 
\hat{\cal Z}(t_{f}, \theta_{f}; t_{i}, \theta_{i})
\eeq
from the initial time \mb{t_{i}} and \mb{\theta_{i}=0} 
to the final time \mb{t_{f}} and \mb{\theta_{f}=0}. Obviously we have
\mb{\hat{\cal Z}(t,\theta)}\mb{=}\mb{\hat{\cal Z}(t,\theta;t_i,\theta_{i})}.

\noi
We now transform all superfield operators 
$\hat{A}(\hat{\zzzz}(t,\theta),\theta)$
into a new representation, which is determined by the external sources:
\beq
\hat{A}'(t,\theta) ~\equiv~ 
\hat{\cal Z}^{-1}(t,\theta)
~\hat{A}(t,\theta)~\hat{\cal Z}(t,\theta) ~.
\eeq
In particular, the original phase space operators turn into
\beq
\hat{\zzzz}'^{A}(t,\theta)~=~\hat{\cal Z}^{-1}(t,\theta)
~\hat{\zzzz}(t,\theta)~\hat{\cal Z}(t,\theta)~.
\eeq 
An important property is that, due to the boundary condition (\ref{zbc}),
all operators coincide in their two different representations at the initial
point $t_i$ and \mb{\theta_{i} = 0}:
\beq
    \hat{A}'(t_{i},\theta_{i})  ~=~ \hat{A}(t_{i},\theta_{i}) ~.
\eeq

\noi
{}From the original Heisenberg equation of motion \e{sh} we find, by
substitution, that the new equation of motion reads slightly different:
\beq
i\hbar D\hat{\zzzz}'^{A}(t,\theta) 
~=~ -\left[\hat{Q}'_{\hat{\Psi}}(t,\theta)
 +  \Delta\hat{Q}'(t,\theta),~\hat{\zzzz}'^{A}(t,\theta) \right]~,
\label{shprime}
\eeq
or for a general operator
\beq
i\hbar D \hat{A}'(t,\theta)~=~-\left[\hat{Q}'_{\hat{\Psi}}(t,\theta)
 +  \Delta\hat{Q}'(t,\theta),~ \hat{A}'(t,\theta) \right]
 +i\hbar {\cal D} \hat{A}'(t,\theta)~. \label{genprime}
\eeq
This precisely describes the dynamics of a shifted 
\mb{\hat{Q}_{\hat{\Psi}}}-operator
\mb{\hat{Q}'_{\hat{\Psi}}+  \Delta\hat{Q}'(t,\theta)}
that is, one with the superfields $J_A$ and $\zzzz^{*}$ playing just
those r\^{o}les of sources as we described above.

\noi
Note that the fundamental supercommutation relations are unaltered by
switching to the new representation:
\bea
[\hat{Q}'_{\hat{\Psi}}(t,\theta),
\hat{Q}'_{\hat{\Psi}}(t,\theta)] &=& 0  \cr
[\hat{\Omega}'_{\hat{\Psi}}(t,\theta),
\hat{\Omega}'_{\hat{\Psi}}(t,\theta)] &=& 0  \cr
[\hat{H}'_{\hat{\Psi}}(t,\theta),
\hat{\Omega}'_{\hat{\Psi}}(t,\theta)] &=& 0 ~.
\eea
But because of the $t$ and $\theta$-dependent sources, the transformed BRST
operator $\hat{\Omega}'_{\hat{\Psi}_{0}}$ is no longer a constant of 
motion. Indeed, from \e{genprime}
\beq
i\hbar D\hat{\Omega}'_{\hat{\Psi}_{0}}(t,\theta) ~=~ 
(-1)^{\epsilon_{A}+1}J_A(t,\theta)[\hat{\zzzz}'^A(t,\theta),
\hat{\Omega}'_{\hat{\Psi}_{0}}(t,\theta)] ~. \label{qprimeeq}
\eeq

\noi
Starting from \eq{zsol}, we now obtain the following identity:
\bea
\dedel{\zzzz^{*}_{A}(t, \theta)}
\hat{\cal Z}(t_{f},\theta_{f} ; t_{i} , \theta_{i})
&~=~&\dedel{\zzzz^{*}_{A}(t, \theta)}~
T \exp\left[  (i\hbar)^{-1} I~ \Delta\hat{Q} \right] (t_{f},\theta_{f})\cr
&~=~&- (i\hbar)^{-2} \hat{\cal Z}(t_{f},\theta_{f};t,\theta)~
K(t_{f},\theta_{f} ; t , \theta)~
[\hat{\zzzz}^{A}(t,\theta),\hat{\Omega}_{\hat{\Psi}_{0}}(t,\theta)] 
~\hat{\cal Z}(t,\theta; t_{i} , \theta_{i}) \cr
&~=~&- (i\hbar)^{-2}\hat{\cal Z}(t_{f},\theta_{f} ; t_{i} , \theta_{i})~
[\hat{\zzzz}'^{A}(t,\theta),\hat{\Omega}'_{\hat{\Psi_{0}}}(t, \theta)] ~,
\eea
where we used that \mb{K(t_{f},\theta_{f} ; t , \theta)=1},
because \mb{\theta_{f}=0}.
This in turn allows us to rewrite the evolution equation \e{qprimeeq}
for $\hat{\Omega}'_{\hat{\Psi}_{0}}$ in the form of
\beq
 (-1)^{\epsilon_{A}+1}  J_A(t,\theta)
\dedel{\zzzz^{*}_{A}(t, \theta)}
 \hat{\cal Z} (t_{f},\theta_{f} ; t_{i} , \theta_{i}) 
~=~-(i\hbar)^{-1}\hat{\cal Z}(t_{f},\theta_{f} ; t_{i} , \theta_{i})~
 D\hat{\Omega}'_{\hat{\Psi}_{0}}(t,\theta)  ~.\label{qagaineq}
\eeq
We let $|0\rangle_{\hat{\Psi}_{0}}$ denote the vacuum state; it also
satisfies the physical-state condition \e{phys}.
We can define the generator ${\cal Z}(J,\zzzz^*)$
or action $W_{c}$ for connected diagrams as
\beq
{\cal Z}(J,\zzzz^*)~=~
\exp\left[\frac{i}{\hbar}W_{c}(J,\zzzz^*)\right] ~=~ 
{}_{\hat{\Psi}_{0}}\!\langle 0|\hat{\cal Z}
(J,\zzzz^*; t_{f},\theta_{f} ; t_{i} , \theta_{i})
|0 \rangle_{\hat{\Psi}_{0}} ~.
\eeq

\noi
We can now finally derive the Ward identity in very compact form. {}From 
\eq{qagaineq} we get, upon integrating over $t$ from $t_{i}$ to $t_{f}$,
and over $\theta$ as well:
\beq
\int_{t_{i}}^{t_{f}} \!  J_{A}(t,\theta)~ dt~ d\theta~
\dedel{\zzzz^{*}_{A}(t,\theta)} W_{c}(J,\zzzz^{*}) ~=~ 0 ~.\label{ward}
\eeq
Although this relation appears to be derivable directly from the equations
of motion only, it encodes in an essential way the gauge (BRST) symmetry
of the system. This is clearly seen when one traces it back to the evolution
equation (\ref{qprimeeq}).
The equation (\ref{ward}) can also be used to derive a superfield analog of 
the Master Equation for the effective action. We shall return to this in 
section 4.

\setcounter{equation}{0}
\section{A Super Phase-Space Path Integral}
\label{secpathint}

\noi
We proceed from the operator formalism to the naive (quasiclassical)
phase space path integral in the usual manner. 
First we introduce the graded super Poisson
bracket
\beq
\{F(\zzzz(t,\theta)),G(\zzzz(t,\theta))\} 
~\equiv~ F\lpart_{A}\omega^{AB}\rpart_{B} G ~.\label{pb}
\eeq
We next consider the classical counterparts of all the previous operator
relations. For $\Omega(\zzzz(t,\theta))$ and $H(\zzzz(t,\theta))$,
\beq
\{\Omega(\zzzz(t,\theta)),\Omega(\zzzz(t,\theta))\} ~=~ 0
 ~~~~~~~~{\mbox{\rm and}}~~~~~~~~~
 \{H(\zzzz(t,\theta)),\Omega(\zzzz(t,\theta))\}~=~ 0~.
\label{homegapb}
\eeq
The Grassmann-odd $Q(\zzzz)$ and $\bar{Q}(\zzzz)$ are defined by
\begin{eqnarray}
Q(\zzzz(t,\theta),\theta) &~\equiv~& \Omega(\zzzz(t,\theta)) + 
\theta H(\zzzz(t,\theta)) ~, \cr
\bar{Q}(\zzzz(t,\theta),\theta) &~\equiv~& \Omega(\zzzz(t,\theta)) - 
\theta H(\zzzz(t,\theta)) ~.\label{qdefpb}
\end{eqnarray}
They are nilpotent in terms of the Poisson bracket, by virtue of eq. 
(\ref{homegapb}):
\beq
\{Q(\zzzz(t,\theta)),Q(\zzzz(t,\theta))\} ~=~ 
\{\bar{Q}(\zzzz(t,\theta)),\bar{Q}(\zzzz(t,\theta))\} ~=~
\{Q(\zzzz(t,\theta)),\bar{Q}(\zzzz(t,\theta))\} ~=~ 0 ~.\label{qpb}
\eeq

\noi
The classical equation of motion is taken to be
\beq
D\zzzz^{A}(t,\theta) 
~=~ -\{Q_{\Psi}(\zzzz(t,\theta),\theta),\zzzz^{A}(t,\theta)\} ~.
\label{shpb}
\eeq
By the same mechanism as before, this is equivalent to
\bea
\frac{d}{dt}\zzzz^A(t,\theta) 
&~=~& -\{H_{\Psi}(\zzzz(t,\theta)),\zzzz^A(t,\theta)\} 
\label{sh1pb}\\
\frac{d}{d\theta}\zzzz^A(t,\theta) &~=~& 
-\{\Omega_{\Psi}(\zzzz(t,\theta)),\zzzz^A(t,\theta)\} ~. 
\label{shomegapb}
\eea

\noi
Going through the analogous manipulations as in the operator formulation,
we find, as expected, the following relations:
\bea
\frac{d}{d\theta}H_{\Psi}(\zzzz(t,\theta)) &=& 
-\{\Omega_{\Psi}(\zzzz(t,\theta)),
H_{\Psi}(\zzzz(t,\theta))\} ~=~ 0 \cr 
\frac{d}{d\theta} \Omega_{\Psi}(\zzzz(t,\theta)) &=& 
-\{\Omega_{\Psi}(\zzzz(t,\theta)),
\Omega_{\Psi}(\zzzz(t,\theta))\} ~=~ 0 \cr
\frac{d}{dt}H_{\Psi}(\zzzz(t,\theta))
 &=& -\{H_{\Psi}(\zzzz(t,\theta)),
H_{\Psi}(\zzzz(t,\theta))\} ~=~ 0 \cr
\frac{d}{dt}\Omega_{\Psi}(\zzzz(t,\theta)) &=& -
\{H_{\Psi}(\zzzz(t,\theta)),
\Omega_{\Psi}(\zzzz(t,\theta))\} ~=~ 0 ~.
\eea
Thus $H_{\Psi}(\zzzz(t,\theta))$ and $\Omega_{\Psi}(\zzzz(t,\theta))$ 
are constants of motion in terms of 
evolution in both $t$ and $\theta$.

\noi
The two superfield equations of motion (\ref{sh1}) and (\ref{shomega})
are therefore equivalent to
\bea
\dot{\zzzz}_{0}^{A}(t)
 &=& -\{H_{\Psi}(\zzzz_{0}(t)),\zzzz_{0}^{A}(t)\}
\label{g0eqm} \\
\zzzz_{1}^{A}(t) &=&
 -\{\Omega_{\Psi}(\zzzz_{0}(t)),\zzzz_{0}^{A}(t)\} ~,\label{g1eqm}
\eea
completely in analogy with the operator relations.

\vspace{0.2cm}
\subsection{\sc The Action}
\label{secaction}

The next step consists in proposing an action by means of which the
above superfield equation of motion, \eq{shpb}, will follow by a 
variational principle. Consider\footnote{
Here we make use of Berezin integration
\mb{\int d\theta \equiv \frac{d}{d\theta}}.
To obtain an action with ghost number $0$ one could divide the left hand side
of \e{susyaction} by $\eta$ and use the prescriptions given in 
Section \ref{secghostnumber}. 
For simplicity we prefer the present formulation.}
\bea
S(\zzzz) &=& \int\! dt~ d\theta \left[
\Hf \zzzz^{A}(t,\theta) ~\omega_{AB}~
D\zzzz^{B}(t,\theta)(-1)^{\epsilon_{B}} 
- Q_{\Psi}(\zzzz(t,\theta),\theta)\right] \cr
&=& - \int\! dt ~d\theta \left[
\Hf(D\zzzz^{A}(t,\theta))~ \omega_{AB}~\zzzz^{B}(t,\theta)
+ Q_{\Psi}(\zzzz(t,\theta),\theta)\right] ~,
\label{susyaction}
\eea
where $\omega_{AB}$ is the inverse of $\omega^{AB}$. By variation we
precisely obtain the equation of motion (\ref{shpb}). The above action $S$
is therefore a good candidate to be exponentiated, and integrated over
in the superfield path integral:
\beq
{\cal Z} ~=~ \int [d\zzzz] \  \exp\left[\frac{i}{\hbar}S(\zzzz)\right] ~.
\label{susypathint}
\eeq
The functional superfield integration over 
$\zzzz^{A}(t,\theta)$ contains no additional measure factor. In a proper
treatment of the phase-space path integral, one should start with the
multiplication algebra of ``symbols'' \cite{BF}, and a path integral
defined through a precise limiting procedure from a suitable 
discretization,
for which eq. (\ref{susypathint}) is just the formal counterpart. For
simple regulators in quantum field theory, the above definition may suffice,
at least as far as the perturbative expansion is concerned.

\noi
It is a remarkable fact, already hinted at in the operator
formulation, that the BRST charge $\Omega$ and the Hamiltonian $H$
enter on an almost equal footing. In the superspace action (\ref{susyaction})
we again see that it is $Q$, the ``superfield combination'' of these
two fundamental objects, which plays the r\^{o}le of the superfield
Hamiltonian.

\noi
For consistency we require that the above path integral reduces
to the usual phase space path integral upon integration over 
$\theta$ in the action, and upon functional integration over the
superfield components $\zzzz_{1}^{A}(t)$. The $\theta$-integration in
the action is straightforward, as it simply projects out the $\theta$-term
in the integrand. We get:
\beq
S(\zzzz_{0},\zzzz_{1}) ~=~ \int\! dt \left[\Hf\zzzz_{0}^{A}(t)
~\omega_{AB}~\dot{\zzzz}_{0}^{B}(t) 
+ \frac{1}{2}(-1)^{\epsilon_B}\zzzz_{1}^{A}(t) ~\omega_{AB}~\zzzz_{1}^{B}(t) 
 - H_{\Psi}(\zzzz_{0}(t) )
-\zzzz_{1}^{A}(t)~\partial_{A}\Omega_{\Psi}(\zzzz_{0}(t))\right] ~.
\eeq
As an intermediate check, we note that the action in this form implies
the equations of motion (\ref{g0eqm}) and (\ref{g1eqm}) 
for $\zzzz_{0}^{A}(t)$ and $\zzzz_{1}^{A}(t)$, respectively. 
Let us next perform the $\zzzz_{1}$-integration.
After completing the square, and making use of the nilpotency condition
for $\Omega_{\Psi}$ in \eq{homegapb}, we find 
\beq
{\cal Z} ~=~ \int [d\zzzz_{0}] ~{\rm Pf}(\omega)~\exp\left[\frac{i}{\hbar}
\int\! dt \left\{\frac{1}{2}\zzzz_{0}^{A}(t)~\omega_{AB}~
\dot{\zzzz}_{0}^{B}(t) - H_{\Psi}(\zzzz_{0}(t))\right\}\right] ~,
\label{reducedpathint}
\eeq
which precisely is the required expression. Even the canonically
invariant Liouville measure is correctly reproduced -- as a result of
the gaussian $\zzzz_{1}$-integration. In fact, the fields $\zzzz_{1}(t)$ 
play two r\^{o}les in the phase-space path integral. On-shell, at the
classical level, they are identitied with the BRST-transformed
$\zzzz^A_0(t)$-fields, while off-shell, after a shift, they can be 
viewed as Pfaffian ghosts.

\vspace{0.2cm}
\subsection{\sc Gauge Independence in the Superfield Formulation}
\label{secgaugeindsuper}

The BFV Theorem \cite{BFV} provides the path integral analog of
the operator proof of gauge independence.
Gauge independence can be proven in both the reduced path integral
\e{reducedpathint} (where $\zzzz_{1}$ is integrated out), and in the
full superfield path integral. We start with the superfield version 
because it is simpler than the component version.  
One might have expected that the natural object to replace
the BRST operator $\Omega_{\Psi_{0}}$ in the required transformation
of variables would be the nilpotent ``super-Hamiltonian'' $Q_{\Psi}$.
This is, however, not the case, as we shall see below. While it is
possible to prove gauge independence by such a replacement, a few
other complications occur. For this reason we have decided to relegate
this type of proof to an appendix, while we here prove the 
superfield analog of the BFV Theorem by means of transformations
based on $\Omega_{\Psi_{0}}$. It is of course still entirely based
on the superfields $\zzzz^A$, with no need to split them up into
components.

\noi
Let us adapt the following natural conventions: 
For a general function \mb{F=F(\zzzz(t,\theta),\theta)}
with explicit $\theta$-dependence define
\bea
F_{0}(\zzzz(t,\theta))&=&F(\zzzz(t,\theta),0)~, \cr
F_{1}(\zzzz(t,\theta))&=&{{\partial}\over {\partial\theta}} 
 F(\zzzz(t,\theta),\theta)~,
\eea
so that \mb{F=F_{0}+\theta F_{1}}.

\noi
Consider the superfield path integral \e{susypathint} with
exponential type of gauge fixing functions 
\mb{\Psi=\Psi'}\mb{=}\mb{\Psi'_{0}+ \theta\Psi'_{1}}.
Let us assume that the gauge fixing functions have been changed
infinitesimally \mb{\Psi' \to \Psi'+ \delta\Psi'} 
causing the Hamiltonian part 
\mb{Q_{\Psi}~=~ e^{{\rm ad}\Psi'}Q} of the action to 
change with
\bea
 \delta Q_{\Psi}&=& \left\{ \EE(\delta\Psi') ,~ Q_{\Psi} \right\} \cr
&=&  \left\{(\EE(\delta\Psi'))_{0} ,~ Q_{\Psi} \right\} 
+ \theta  \left\{(\EE(\delta\Psi'))_{1} ,~ \Omega_{\Psi_{0}} \right\} ~,
\label{externvar}
\eea
where 
\beq
\EE(F) ~\equiv~ \int_{0}^{1}\! d\alpha~ 
e^{+\alpha {\rm ad}\Psi'} F ~,\label{eepbdef}
\eeq
and
\beq
{\rm ad} F ~\equiv~ \{F,~\cdot~\} ~.
\eeq
We will now show that this change is cancelled through 
an internal rearrangement of the path integral, and we can hence conclude that
the path integral does not depend on the chosen gauge fixing.
The first term in \e{externvar} is cancelled by a canonical
transformation 
\beq
\delta\zzzz^{A}(t,\theta)~=~
\{ \zzzz^{A}(t,\theta), (\EE(\delta \Psi'))_{0}(\zzzz(t,\theta)) \}~.
\eeq
A canonical transformation produces no Jacobian, and
the kinetic term is invariant (up to a boundary term)
because there is -- by construction -- no explicit $\theta$-dependence inside
\mb{(\EE(\delta\Psi'))_{0} (\zzzz(t,\theta))}. The change in the 
Hamiltonian part yields precisely
\beq
 \left\{ Q_{\Psi}  ,~(\EE(\delta\Psi'))_{0}  \right\}~,
\eeq
\ie minus the first term in \e{externvar}.
The second term in \e{externvar} is cancelled by a BRST type of 
transformation 
\beq
 \delta\zzzz^{A}(t,\theta) ~=~
\{ \zzzz^{A}(t,\theta),~\Omega_{\Psi_{0}}(\zzzz(t,\theta)) \}~ \mu~,
\eeq
where $\mu$ is an odd functional of the form
\beq
   \mu~=~- \Ih \int \! dt' ~d\theta'~ 
\theta'~ (\EE(\delta \Psi'))_{1}(\zzzz(t',\theta'))~.
\eeq
Here both the Hamiltonian and the kinetic part of the action is invariant.
The latter because there is no explicit $\theta$-dependence in
\mb{\Omega_{\Psi_{0}}}. Finally the Jacobian produces
the second term:
\bea
  J-1&=& (-1)^{\epsilon_{A}} \int \! \int \!
  \delta \zzzz^{A}(t,\theta) \deder{\zzzz^{A}(t',\theta')}~
  dt' ~d\theta'~  \delta(t'-t)~ \delta(\theta'-\theta)~ dt ~d\theta  \cr
&=& \Ih \int \!  dt ~d\theta~ \theta~
 \left\{ (\EE(\delta\Psi'))_{1}(\zzzz(t,\theta)) ,~ 
\Omega_{\Psi_{0}}(\zzzz(t,\theta)) \right\}~.
\eea 

\noi
We have thus shown that the change in gauge, $\Psi' \to \Psi' + \delta\Psi'$,
can be reabsorbed by a combination of a superfield canonical 
transformation and a superfield BRST transformation. The path integral is
hence formally independent of $\Psi'$.

\vspace{0.2cm}
\subsection{\sc Gauge Independence in the Original Sector}

We now outline the proof in the formulation \e{reducedpathint} 
where $\zzzz_1$ has been integrated out. 
It is interesting it is own right to see how the proof generalizes
in the original sector. 
Because of the ``exponential'' gauge fixing 
\mb{\Psi=\Psi'=\Psi'_{0}+ \theta\Psi'_{1}}, 
the proof differs in some details from the original proof \cite{BFV}.

\noi
We first introduce, in analogy with \eq{edef},
\beq
\Eo(F) ~\equiv~ \int_{0}^{1}\! d\alpha~ 
e^{+\alpha {\rm ad}\Psi_{0}}F ~.\label{epbdef}
\eeq
An infinitesimal change in the gauge fixing
produces the following change in the Hamiltonian
\beq
 \delta H_{\Psi}~=~\left\{   \delta\Xi_{0} ,  H_{\Psi} \right\}
+\left\{ \delta\Xi_{1} , \Omega_{\Psi_{0}} \right\}\label{hamchange}
\eeq
where
\bea
   \delta\Xi_{0} &\equiv&\Eo(\delta \Psi_{0})~, \cr
  \delta\Xi_{1} &\equiv&\Eo(\delta \Psi'_{1})
+\left\{\Eo( \Psi'_{1}) , \Eo(\delta \Psi_{0})  \right\}
+\left \{ \int_0^1\! d\alpha \int_0^1\! d\beta \left( 
e^{+\alpha\beta {\rm ad}\Psi_{0}}\delta \Psi_{0} \right)  ,~
e^{+{\rm ad}\Psi_{0}} \Psi'_{1}   \right\}  ~.
\eea
Now the idea is the same as in the superfield approach.
First, we perform a canonical transformation 
\beq
 \delta\zzzz^{A}_{0}(t)
~=~\{ \zzzz^{A}_{0}(t), \delta\Xi_{0}(\zzzz_{0}(t)) \}~.
\eeq 
This changes the Hamiltonian part with
\beq
\delta\int \! dt~ H_{\Psi}(\zzzz_{0}(t))
~=~ \int \! dt \{ H_{\Psi}(\zzzz_{0}(t)), \delta\Xi_{0}(\zzzz_{0}(t)) \}~.
\eeq
Next, consider the BRST variation
\beq
 \delta\zzzz^{A}_{0}(t)~=~ 
\{ \zzzz^{A}_{0}(t),\Omega_{\Psi_{0}}(\zzzz_{0}(t)) \} ~\mu~.
\eeq
with
\beq
 \mu~=~ - \frac{i}{\hbar} \int \! dt'~ \delta\Xi_{1}(\zzzz_{0}(t'))~.
\eeq
The action is invariant, but the Jacobian equals
\beq
J-1~=~(-1)^{\epsilon_{A}} \int \! \int \!
  \delta \zzzz_{0}^{A}(t) \deder{\zzzz_{0}^{A}(t')}~ dt' ~\delta(t'-t)~ dt
~=~ \Ih \int \!  dt~  \left\{ \delta\Xi_{1}(\zzzz_{0}(t)) ,~ 
\Omega_{\Psi_{0}}(\zzzz_{0}(t)) \right\}~.
\eeq
The two new terms in \eq{hamchange} are thus explicitly cancelled,
and we conclude again that the path integral is independent of $\Psi'$.


\setcounter{equation}{0}

\section{Super Antifield Formalism}

So far our considerations have been restricted to the Hamiltonian
counterpart of BRST quantization. It is natural to seek an extension
of this to the Lagrangian framework, which, for example, with little
effort provides a manifestly Lorentz invariant description. To encompass
the complete set of all possible gauge theories, we know that the
appropriate language should be that of the field-antifield formalism
\cite{BV}. This leads us to consider the reformulation of this
field-antifield construction in superfield language.

\noi
The essential ingredient in the field-antifield formalism is a
Grassmann-odd and nilpotent operator $\Delta$, whose
failure to act like a differentiation defines a Grassmann-odd bracket,
the antibracket, and an associated antisymplectic geometry. Before
we proceed with the derivation of a superfield analog of the 
field-antifield formalism, it is therefore useful to first pause and
consider an appropriate generalization of these concepts to superfields.
Although we will not need it in the present preliminary stage, we choose
do it in the more general covariant formulation, where the antisymplectic
coordinates have not necessarily been specified in Darboux form. It will
be useful for
later developments, where one would like to build a more abstract and
coordinate-independent field-antifield formalism in superfield language.

\vspace{0.2cm}
\subsection{\sc General Covariant Theory}
\label{secantifield}

In this subsection we therefore descibe a general odd symplectic superspace 
with $4N$ variables 
\beq
\Gammm^{A}(t,\theta) 
~=~ (\Phi^{\alpha}(t,\theta); \Phi^{*}_{\alpha}(t,\theta))~,
\eeq
where \mb{\Phi^{\alpha}(t,\theta)} and \mb{\Phi^{*}_{\alpha}(t,\theta)}
have the {\em same} statistics \mb{\epsilon_{\alpha}}.
We shall later, in section \ref{sechammastereq}, let 
our previous superfield phase-space variables $z^A(t,\theta)$
play the r\^{o}les of $\Phi^{\alpha}(t,\theta)$.
In that particular case $N$ is thus even. 

\noi
Given a volume density $\rho$ and a Grassmann-odd symplectic metric  
\mb{E^{AB}(t,\theta;t',\theta')} the covariant ``odd superfield Laplacian'' 
$\Delta$ reads
\beq
  \Delta ~=~ {1 \over 2}(-1)^{\epsilon_{A}} \rho^{-1}
 \int \! \int \! dt~d\theta~  \dedel{\Gammm^{A}(t,\theta)}
~\rho~ E^{AB}(t,\theta;t',\theta')~dt'~d\theta'~ 
\dedel{\Gammm^{B}(t',\theta')}~.
\eeq
We assume that the the measure density and the antisymplectic structure are 
compatible in the sense that $\Delta$ is nilpotent: $\Delta^2 \!=\! 0$.
This $\Delta$-operator gives rise to a superfield antibracket in the
conventional manner, through its failure to act as a derivation:
\bea
  (F,G) &=& (-1)^{\epsilon_{F}}
 [[\stackrel{\rightarrow}{\Delta},F],G]1
~=~ - (-1)^{(\epsilon_{F}+1)(\epsilon_{G}+1)} (G,F) \cr
&=&  \int \! \int \! F\deder{\Gammm^{A}(t,\theta)}
~dt~d\theta~ E^{AB}(t,\theta;t',\theta')~dt'~d\theta'~ 
\dedel{\Gammm^{B}(t',\theta')} G ~.
\eea
We have here also indicated its symmetry property.

\noi
The antisymplectic metric should be non-degenerate, it should 
have Grassmann parity
\beq
\epsilon(E^{AB}(t,\theta;t',\theta'))
 ~=~ \epsilon_{A}+ \epsilon_{B} + 1~,
\label{egrass}
\eeq
and symmetry
\beq
E^{BA}(t',\theta';t,\theta)
 ~=~ -(-1)^{(\epsilon_{A}+1)(\epsilon_{B}+1)} 
 E^{AB}(t,\theta;t',\theta')~. 
\eeq
It satisfies the Jacobi identity
\beq
\sum_{{\rm cycl.}~A,B,C} (-1)^{(\epsilon_{A}+1)(\epsilon_{C}+1)} 
 \int E^{AD}(t^{1},\theta^{1};t^{4},\theta^{4}) ~dt^{4}~ d\theta^{4}  
\dedel{\Gamma^{D}(t^{4},\theta^{4})}
E^{BC}(t^{2},\theta^{2};t^{3},\theta^{3})~=~0~.
\eeq

\noi
For a general bosonic vector field
\bea
  X&=& \int \! X^{A}(t,\theta)~dt~ d\theta~ \dedel{\Gammm^{A}(t,\theta)} \cr
   &=& \int \! X_{0}^{A}(t)~dt~ \dedel{\Gammm_{0}^{A}(t)}
       +\int \! X_{1}^{A}(t)~dt~ \dedel{\Gammm_{1}^{A}(t)}~,
\eea
with components \mb{X^{A}(t,\theta)=X_{0}^{A}(t)+\theta X_{1}^{A}(t)}
of Grassmann parity \mb{\epsilon(X^{A})=\epsilon_{A}}, the divergence
\mb{{\rm div}_{\rho} X} is defined as
the proportionality factor between the Lie-derivative of the 
volume-form \mb{{\cal L}_{X} \vol} and the volume-form  \mb{\vol} itself:
\beq
  {\cal L}_{X} \vol~=~{\rm div}_{\rho} X~ \vol~,
\eeq
or 
\bea
{\rm div}_{\rho} X &=& (-1)^{\epsilon_{A}} \rho^{-1}  \int dt~d\theta~  
\dedel{\Gammm^{A}(t,\theta)} ~\rho~ X^{A}(t,\theta) \cr
 &=& (-1)^{\epsilon_{A}} \rho^{-1}  \int dt~  
\dedel{\Gammm_{0}^{A}(t)} ~\rho~ X_{0}^{A}(t)
+(-1)^{\epsilon_{A}+1} \rho^{-1}  \int dt~  
\dedel{\Gammm_{1}^{A}(t)} ~\rho~ X_{1}^{A}(t)
\eea

\noi
Note that if \mb{X^{A}(t,\theta)=X^{A}(\Gamma(t,\theta),t,\theta)} is
an ``ultra-local vector field'', then
\beq
{\rm div}_{\rho} X = \rho^{-1}X[ \rho ]~.
\eeq
If we restrict ourselves to a class of coordinate patches
that are pairwise mutually connected by ultra-local super transformations
\mb{\Gamma'^{A} = F^A(\Gamma(t,\theta),t,\theta)}, then it is consistent
to choose \mb{\rho} to be the same for all patches, \ie a 
coordinate change within  
the above mentioned class will not change the value of $\rho$.
In other words, under this restricted class of reparametrizations
$\rho$ behaves not only as a scalar density, but as a scalar.
{}From a geometrical point of view there is hence no reason to insert a 
non-trivial
measure density $\rho$ different from $1$. 
When  \mb{\rho=1} the divergence of an ultra-local vector field 
\mb{X^{A}(t,\theta)=X^{A}(\Gamma(t,\theta),t,\theta)}
vanishes identically, so ultra-locally we have an analogue of the
Liouville Theorem.
For a general Hamiltonian (but not necesseraly ultra-local) 
vectorfield \mb{X_{F}=(F,\cdot)=-(\cdot,F)}, 
where $F$ is Grassmann-odd function, the divergence  
\beq
{\rm div}_{\rho} X_{F}~=~ - 2 \Delta(F)  
\eeq
is given by the odd Laplacian, as in the normal case \cite{KN}.

\vspace{0.2cm}
\subsection{\sc Darboux Coordinates}
\label{secdarboux}

In Darboux coordinates the fundamental antibracket relations read
in superfield form:
\beq
E^{\alpha} {}^{*}_{\beta}(t,\theta;t',\theta')
~=~(\Phi^{\alpha}(t,\theta),\Phi_{\beta}^{*}(t',\theta')) 
~=~  \delta^{\alpha}_{\beta} ~\delta(t-t')~ \delta(\theta-\theta')~.
\eeq
In components the non-vanishing bracket relations become
\bea
   (\Phi_{0}^{\alpha}(t),\Phi_{1\beta}^{*}(t'))
 &=& (-1)^{\epsilon_{\alpha}} \delta^{\alpha}_{\beta} ~\delta(t-t')~, \cr
   (\Phi_{1}^{\alpha}(t),\Phi_{0\beta}^{*}(t'))
 &=&  \delta^{\alpha}_{\beta} ~\delta(t-t')~.
\eea
We see that the \mb{(-1)^{\epsilon_{\alpha}}\Phi_{1\alpha}^{*}(t)}  
are the antisymplectic conjugate variables to the original variables 
\mb{\Phi_{0}^{\alpha}(t)}.
In the same manner \mb{\Phi_{0\alpha}^{*}(t)} are conjugate 
to the superpartners \mb{\Phi_{1}^{\alpha}(t)}. 
\beq
  (F,G) ~=~  \int F \left(
\deder{\Phi^{\alpha}(t,\theta)}~dt~ d\theta ~
 \dedel{\Phi_{\alpha}^{*}(t,\theta)}-(-1)^{\epsilon_{\alpha}}
\deder{\Phi_{\alpha}^{*}(t,\theta)} ~dt~ d\theta ~
 \dedel{\Phi^{\alpha}(t,\theta)} \right) G ~.
\eeq
If \mb{\rho=1} the $\Delta$-operator reduces to
\bea
\Delta &=&- \int  \dedel{\Phi^{\alpha}(t,\theta)} dt ~d\theta~
\dedel{\Phi_{\alpha}^{*}(t,\theta)}
~=~(-1)^{\epsilon_{\alpha}} \int  dt ~d\theta~
\dedel{\Phi^{\alpha}(t,\theta)} 
\dedel{\Phi_{\alpha}^{*}(t,\theta)} \cr
&=&\int \! dt \left(\dedel{\Phi_{0}^{\alpha}(t)}
 \dedel{\Phi_{1\alpha}^{*}(t)}
-(-1)^{\epsilon_{\alpha}} \dedel{\Phi_{1}^{\alpha}(t)}
 \dedel{\Phi_{0\alpha}^{*}(t)} \right)~.
\eea

\vspace{0.2cm}
\subsection{\sc Path Phase-Space Antibracket}

As an aside, we note that one
can lift \cite{bering} the equal-$t$--equal-$\theta$ 
even symplectic structure 
\beq
\{F(\zzzz(t,\theta)),G(\zzzz(t,\theta))\} 
~\equiv~ F\lpart_{A}~\omega^{AB}(\zzzz(t,\theta))
~\rpart_{B} G ~.\label{pbequal} 
\eeq
to an odd path-space symplectic structure
\beq
(F,G)~=~  \int \int \! F\deder{\zzzz^{A}(t,\theta)}
~dt~d\theta~ E^{AB}(t,\theta;t',\theta')~dt'~d\theta'~ 
\dedel{\zzzz^{B}(t',\theta')} G~.\label{oddpb} 
\eeq
by the ultralocal ansatz
\beq
  E^{AB}(t,\theta;t',\theta')~=~\omega^{AB}(\zzzz(t,\theta))~
\delta(t-t')~ \delta(\theta-\theta')(-1)^{\epsilon_{B}}~.
\eeq
Note that the Jacobi identity for \mb{\omega^{AB}} carries over to the
Jacobi identity for \mb{E^{AB}}. With the measure density \mb{\rho=1}
we can form a nilpotent odd Laplacian 
\bea
  \Delta &=& {1 \over 2}(-1)^{\epsilon_{A}} 
 \int \! \int \! dt~d\theta~  \dedel{\zzzz^{A}(t,\theta)}
~ E^{AB}(t,\theta;t',\theta')~dt'~d\theta'~ 
\dedel{\zzzz^{B}(t',\theta')} \cr
&=& \int \! dt \left\{ 
(-1)^{\epsilon_{A}} \omega^{AB}(\zzzz_{0}(t))~
\dedel{\zzzz^{B}_{1}(t)}\dedel{\zzzz^{A}_{0}(t)}
\right. \cr && \left.
-{1 \over 2} (-1)^{\epsilon_{A}}
\zzzz^{C}_{1}(t)~\partial_{C}\omega^{AB}(\zzzz_{0}(t))~
\dedel{\zzzz^{B}_{1}(t)}\dedel{\zzzz^{A}_{1}(t)}\right\}~.
\eea
In components the corresponding antibracket reads
\bea
  (\zzzz^{A}_{0}(t),\zzzz^{B}_{0}(t'))&=&0~, \cr
  (\zzzz^{A}_{0}(t),\zzzz^{B}_{1}(t'))&=&\omega^{AB}(\zzzz_{0}(t))
\delta(t-t')~,\cr
  (\zzzz^{A}_{1}(t),\zzzz^{B}_{1}(t'))&=&
\zzzz^{C}_{1}(t)~\partial_{C}\omega^{AB}(\zzzz_{0}(t))\delta(t-t')~.
\eea
Such an antisymplectic structure has previously, in a different context, 
been considered in \cite{nersessian}.

\noi
Finally, a set of Darboux coordinates
\beq
\zzzz^{A}(t,\theta) 
~=~ (q^{\alpha}(t,\theta); p^{*}_{\alpha}(t,\theta))
\eeq
satisfies
\beq
(q^{\alpha}(t,\theta),p_{\beta}^{*}(t',\theta')) 
~=~  \delta^{\alpha}_{\beta} ~\delta(t-t')~ \delta(\theta-\theta')~.
\eeq

\vspace{0.2cm}
\subsection{\sc The Hamiltonian Master Equation}
\label{sechammastereq}

\noi
Once the heuristic phase-space path integral has been established in
superspace form, it is a small step to formally derive from it a 
corresponding Lagrangian field-antifield path integral in superfield
form. There have already previously been some suggestions for such
a superfield formulation of the field-antifield formalism \cite{superBV},
but they have not started with the Hamiltonian phase-space path integral
itself, and, as we shall see, this leads to some differences.

\noi
Before proceeding to the formal derivation of the field-antifield
superfield path integral, let us first return briefly to the Hamiltonian
operator formulation. We have already seen how the fundamental Ward identities
for the generator $W_c$ in the phase space formulation could be compactly
cast in the form of \eq{ward}. Let us now go one step further, and
define classical variables in the usual way
\beq
 \zzzz^{A}_{\rm cl}(t,\theta) 
~\equiv~ \dedel{J_{A}(t,\theta)} W_c(J,\zzzz^*)
~=~ - W_c(J,\zzzz^*) \deder{J_{A}(t,\theta)} ~.
\eeq
Furthermore, we assume as usual that this relation can be inverted 
so that we can express $J_{A}(t,\theta)$ in terms of 
$\zzzz^{A}_{\rm cl}(t,\theta)$. We can then define the effective 
action $\Gameff$ in the standard way of a Legendre transform:
\beq
\Gameff(\zzzz_{\rm cl},\zzzz^{*}) ~\equiv~ W_{c}(J,\zzzz^{*})
- \int \! J_{A}(t,\theta)~ dt~ d\theta~ \zzzz^{A}_{\rm cl}(t,\theta) ~,
\eeq
where on the right hand side we insert the solution $J_{A}(t,\theta)$ 
as a function of $\zzzz^{A}_{\rm cl}(t,\theta)$. It follows that 
\beq
 J_{A}(t,\theta)
~=~  \dedel{ \zzzz^{A}_{\rm cl} (t,\theta)} 
\Gameff(\zzzz_{\rm cl},\zzzz^{*})
~=~ - \Gameff(\zzzz_{\rm cl},\zzzz^{*}) 
 \deder{\zzzz^{A}_{\rm cl} (t,\theta)}~.
\eeq
So the Ward identity \e{ward} turns into a generalized Zinn-Justin equation:
\bea
0&=&\int \!
 \Gameff(\zzzz_{\rm cl},\zzzz^{*})~ 
 \deder{\zzzz^{A}_{\rm cl}(t,\theta)}~ dt~  d\theta ~ 
\dedel{\zzzz^{*}_{A}(t,\theta)} ~
\Gameff(\zzzz_{\rm cl},\zzzz^{*}) \cr
  &=& \hf \left( \Gameff(\zzzz_{\rm cl},\zzzz^{*}) , 
\Gameff(\zzzz_{\rm cl},\zzzz^{*}) \right)_{\rm cl} ~.
\label{zinnjustin}
\eea
where we have introduced the antibracket, here with respect to
the ``classical'' superfields $\zzzz_{\rm cl}$ and BRST 
superfield sources $\zzzz^{*}$.
This equation has an uncanny resemblance to only the {\em classical} part
of a conventional Lagrangian Master Equation written in superfield
language. Of course, no approximations are involved at this stage, so the
above equation is exact, as is clear from the way in which it was 
derived directly from exact operator
relations. There are no ``quantum corrections'' to the superfield Master
Equation, when written as above in terms of the effective action.

\noi
One way to derive a more conventional type of Master Equation from the
operator formalism, is to
formally introduce a functional Fourier transform 
$e^{{i \over {\hbar}}W_{H}^{({\rm gf})}(\zzzz,\zzzz^{*})}$
of the generator of connected diagrams ${\cal Z}(J,\zzzz^*)$ 
with respect to the variables  \mb{J \leftrightarrow \zzzz}:
\bea
\exp\left[{i \over {\hbar}} W_{H}^{({\rm gf})}(\zzzz,\zzzz^{*})\right]
&=& \int [dJ] \  \exp\frac{i}{\hbar}\left[ W_{c}(J,\zzzz^{*})
- \int \! J_{A}(t,\theta)~ dt~ d\theta ~  \zzzz^{A}(t,\theta) \right] \cr
\exp\left[{i \over {\hbar}} W_{c}(J,\zzzz^{*})\right]
&=& \int [d\zzzz] \  \exp\frac{i}{\hbar}\left[
W_{H}^{({\rm gf})}(\zzzz,\zzzz^{*})
+ \int \!  J_{A}(t,\theta)~ dt~ d\theta ~ \zzzz^{A}(t,\theta) \right] ~.
\eea
Here $W_{H}^{({\rm gf})}(\zzzz,\zzzz^{*})$ plays the r\^{o}le of an
action for a path integral with field variables $\zzzz^{A}$.
In fact, $W_{H}^{({\rm gf})}(\zzzz,\zzzz^{*})$ is a Hamiltonian counterpart
of a {\em gauge-fixed} Lagrangian BV action. It satisfies a $8N$-dimensional
phase space quantum Master Equation.
To see this, let us write down the phase space odd Laplacian on the doubly
extended space of $\zzzz^A$ and $\zzzz^{*}_{A}$: 
\beq
\Delta_{\zzzz} ~=~(-1)^{\epsilon_{A}}\int \!  dt ~d\theta~ 
\dedel{\zzzz^{A}(t,\theta)} \dedel{\zzzz_{A}^{*}(t,\theta)}~.
\eeq
The Ward identity \e{ward} can now be rewritten as 
a more conventional-looking quantum Master Equation:
\beq
 \Delta_{\zzzz}  \exp\left[\frac{i}{\hbar}
 W_{H}^{({\rm gf})}(\zzzz,\zzzz^{*}) \right] ~=~0~.
\eeq
The $W_{H}^{({\rm gf})}(\zzzz,\zzzz^{*})$ defined in 
the above way does not depend on the gauge fermion $\hat{\Psi}$, and 
it should of course 
be identified with an action already gauge-fixed, \ie no extra
gauge fixing is needed at this point. That is why we added a 
superscript ``$({\rm gf})$''. In particular, one can take the external
sources $\zzzz^*_A$ to vanish.

\vspace{0.2cm}
\subsection{\sc The Lagrangian Master Equation}

Let us now show how analogous results can be 
reproduced from the path integral point of view.
We shall also derive the counterpart where
the {\em some} phase space variables $\Pi_{\alpha}(t,\theta)$
are integrated out. 
\beq
\zzzz^{A}(t,\theta) ~=~ (\Phi^{\alpha}(t,\theta); \Pi_{\alpha}(t,\theta))~.
\eeq
In particular, if one integrates out precisely half of
the phase space variables, taken to be momenta, one obtains the Lagrangian
version of the Master Equation. Our derivation is essentially a
superfield extension of the one presented in ref. \cite{GGT}.

\noi
We start with the Hamiltonian action mentioned in Section \ref{secaction}:
\beq
S^{(0)}(\zzzz) ~=~ \int\! dt~ d\theta \left[
\hf \zzzz^{A}(t,\theta) ~\omega_{AB}~
D\zzzz^{B}(t,\theta)(-1)^{\epsilon_{B}} 
- Q_{\Psi^{(0)}}(\zzzz(t,\theta),\theta)\right] 
\eeq
Here we have included an initial gauge fixing 
\mb{\Psi^{(0)}=\Psi^{(0)}(\zzzz(t,\theta),\theta)} to be 
as general as possible. (Specifically, we shall later need
this initial gauge fixing in the $\Pi_{\alpha}(t,\theta)$
sector). 
The Hamiltonian master action $W_{H}(\zzzz,\zzzz^{*})$ is introduced
as the above action, linearly extended with sources $\zzzz_{A}^{*}$ for the 
BRST transformation. 
\beq
 W_{H}(\zzzz,\zzzz^{*})~=~S^{(0)}(\zzzz) 
+  \int \! \zzzz_{A}^{*}(t,\theta)~ dt~ d\theta ~ 
\{ \zzzz^{A}(t,\theta), \Omega_{\Psi_{0}}(\zzzz(t,\theta)) \}
\eeq
We assume that the remaining gauge fixing is of the linear type
\mb{\Psi = \Psi''_{0}(\Phi(t,\theta)) 
+ \theta\  \Psi''_{1}(\Phi(t,\theta))},
with for simplicity \mb{\Psi''_{0} = 0},
cf.\ discussion in Section \ref{secgaugefix}. 
We choose it so that it does  not depend on the 
momenta $\Pi_{\alpha}(t,\theta)$. 
An exponential type of gauge fixing should be rewritten into a
linear form, or should be included in the initial gauge fixing above.

\noi
We now introduce the gauge fermion functional
\beq
    \psi ~=~  \int \!  dt~ d\theta ~\Psi(\zzzz(t,\theta),\theta)~.
\eeq
Note that this indeed is a fermion. We have the following relation
\beq
   (-1)^{\epsilon_{A}} \dedel{\zzzz^{A}(t,\theta)} \psi
 ~=~ \papal{\zzzz^{A}(t,\theta)} \Psi(\zzzz(t,\theta),\theta)
 ~=~(-1)^{\epsilon_{A}}
  \Psi(\zzzz(t,\theta),\theta) \papar{\zzzz^{A}(t,\theta)}
 ~=~  \psi \deder{\zzzz^{A}(t,\theta)}~.
\eeq
The path integral is given as
\beq
{\cal Z}(J,\zzzz^{*})
~=~ \int [d\zzzz] \  \exp\frac{i}{\hbar}\left[
W_{H}(\zzzz,\zzzz^{*}+\psi\deder{\zzzz})
+ \int \! J_{A}(t,\theta)~ dt~ d\theta ~  \zzzz^{A}(t,\theta) \right] ~.
\eeq
As a first check, note that the Ward identity \e{ward} may be derived
by performing an ordinary BRST variation (with constant $\mu$): 
\beq
   \delta \zzzz^{A}(t,\theta)~=~
\{ \zzzz^{A}(t,\theta),\Omega_{\Psi_{0}}(\zzzz(t,\theta)) \} \mu~.
\eeq

\noi
The phase space Master Equation may independently be derived 
from the fact that $W_{H}(\zzzz,\zzzz^{*})$ is BRST invariant:
\bea
0&=& \int \! \{ \Omega_{\Psi_{0}}(\zzzz(t,\theta)),  \zzzz^{A}(t,\theta)\}
~ dt~ d\theta~ \dedel{ \zzzz^{A}(t,\theta)} 
\exp\frac{i}{\hbar}\left[ W_{H}(\zzzz,\zzzz^{*})\right]  \cr
&=&  \int \! \int \!  dt'~ d\theta'~ \delta(t'-t)~
\delta(\theta'-\theta)~ 
 \{ \Omega_{\Psi_{0}}(\zzzz(t',\theta')),\zzzz^{A}(t',\theta') \}
~dt~ d\theta \cr
&& \times~\dedel{ \zzzz^{A}(t,\theta)} 
\exp\frac{i}{\hbar}\left[ W_{H}(\zzzz,\zzzz^{*})\right]  \cr
&=& -(-1)^{\epsilon_{A}}  \int \! \int \!  dt'~ d\theta'~ \delta(t'-t)~
\delta(\theta'-\theta)~ dt~ d\theta \cr
&& \times~\dedel{ \zzzz^{A}(t,\theta)} 
\left( \{ \zzzz^{A}(t',\theta'),\Omega_{\Psi_{0}}(\zzzz(t',\theta')) \}~
\exp\frac{i}{\hbar}\left[ W_{H}(\zzzz,\zzzz^{*})\right] \right) \cr
&=& -(-1)^{\epsilon_{A}} \int \! \int \!  dt'~ d\theta'~ \delta(t'-t)~
\delta(\theta'-\theta)~ dt~ d\theta \cr
&& \times~\dedel{ \zzzz^{A}(t,\theta)} 
~\dedel{ \zzzz_{A}^{*}(t,\theta)}~
\exp\frac{i}{\hbar}\left[ W_{H}(\zzzz,\zzzz^{*})\right] \cr
&=&- \Delta_{\zzzz}  \exp\left[\frac{i}{\hbar} 
  W_{H}(\zzzz,\zzzz^{*}) \right] ~.\label{phsme}
\eea
Note that the odd Laplacian separates into two pieces:
\beq
  \Delta_{\zzzz}~=~ \Delta_{\Phi}~+~ \Delta_{\Pi}~,
\eeq
each being on Darboux form, cf.\ Section \ref{secdarboux}.

\noi
The idea is now to formally integrate out all $\Pi_{0}$ degrees of freedom.
Although the functional integral may be undoable in closed form, we simply
define the Lagrangian action $W$ through
\beq
\exp\left[\frac{i}{\hbar} W(\Phi,\Phi^{*},\Pi_1,\Pi_1^*,\Pi_0^*)\right]
~=~ \int \! [d\Pi_{0}]~ \exp\left[\frac{i}{\hbar}  
 W_{H}(\zzzz,\zzzz^{*}) \right]~. \label{Wdefi}
\eeq 
This is  manifestly supersymmetric under a supertranslation
\mb{\Pi(\theta) \to \Pi(\theta-\theta_{0})}. The reason why we do
not integrate out the superpartners $\Pi_{1}$ 
is -- as we have seen earlier -- that they are the Pfaffian ghosts.
An integration over $\Pi_{1}$ would in general produce delta functions
in $\Phi_{1}$ so that the action would turn singular.
More precisely, one should require that the action is {\em proper}
\cite{BT1}. As it stands, eq. (\ref{Wdefi}) is the most sensible way
of defining the superfield Lagrangian action $W$ at present. We also
note that the introduction of superfield momentum sources $\Pi_0^*$ and
$\Pi_1^*$ in the Hamiltonian path integral is a choice made by us. It
is not required, but it makes the description more symmetric in phase
space variables, and it simplifies the subsequent derivations.

\noi
Finally, this gives us the Lagrangian Master Equation for $W$:
\bea
\Delta_{\Phi}~\exp\frac{i}{\hbar}\left[ W(\Phi,\Phi^{*},\Pi_1,\Pi_1^*,
\Pi_0^*)\right]
&=& \int \! [d\Pi_{0}]~\Delta_{\Phi}~ \exp\left[\frac{i}{\hbar}  
 W_{H}(\zzzz,\zzzz^{*}) \right] \cr
&=& \int \! [d\Pi_{0}]~(\Delta_{\zzzz}- \Delta_{\Pi} )~ 
\exp\left[\frac{i}{\hbar}   W_{H}(\zzzz,\zzzz^{*}) \right]  ~=~ 0~.
\label{LME}
\eea
where we have used the phase-space Master Equation \e{phsme} 
and the fact that the $\Delta_{\Pi}$-term is a total derivative in $\Pi_{0}$.

\noi
While \mb{\Pi_{1}=\int \! d\theta~ \Pi(\theta)} is a manifest supersymmetric 
variable, the antisymplectic conjugate \mb{\Pi^{*}_{0}} is not.
In a Lagrangian formulation where \mb{\Pi_{0}} is integrated out it
is clear that one should consider \mb{\Pi_{1}} as a auxilary variable
with no antifield attached. Apart from this, there is no difference
in the formal structure between this superfield formulation, and the
usual field-antifield formalism. All relations among antibrackets
and between antibrackets and the $\Delta$-operator have analogous
superfield extensions. Also BRST operators (be they classical or
quantum) can therefore be constructed in the appropriate superfield
form. It may be worthwhile to investigate that issue in further
detail, but this would clearly take us beyond the scope the present paper.


\setcounter{equation}{0}
\section{Conclusion}
\label{secconclus}

\noi
As we have shown in this paper, it is possible to set up a manifestly
BRST-symmetric operator formulation of the quantization of theories
with first-class constraints by means of a straightforward superfield 
extension.  The result has all the features one would have hoped for:
\begin{itemize}
\item The BRST operator $\hat{\Omega}$ and the Hamiltonian $\hat{H}$
enter in a unified manner through the nilpotent operator
\mb{\hat{Q}}\mb{=}\mb{ \hat{\Omega} + \theta\hat{H}}.
\item Each original field (operator) $\hat{\zzzz}^A_{0}(t)$, 
is unified in the corresponding superfield 
\mb{\hat{\zzzz}^{A}(t,\theta)}\mb{=}
\mb{ \hat{\zzzz}^{A}_{0}(t) + \theta\hat{\zzzz}^{A}_{1}(t)} 
with its BRST-transform $\hat{\zzzz}^{A}_{1}(t)$. In the path integral,
this relationship holds on-shell. Both $z_0(t)$ and $z_1(t)$
are integrated over in the path integral, where now $z_1(t)$ act
as Pfaffian ghosts. 
\item The superfield formalism naturally links BRST
transformations with canonical transformations.
\item The operator quantization can be carried through entirely in
the superfield formulation, through exact operator relations at the
superfield level.
\item A phase-space superfield path integral can be set up, which reproduces
the correct equations of motion, and which, upon integrating out the
superfield partners $\zzzz^{A}_{1}$, reduce to the conventional phase
space BFV path integral.
\item A superfield generalization of the BFV Theorem can be proved
through the use of a combination of a BRST transformation with a
canonical transformation. On the subspace of original variables, 
it reduces to a proof of the usual BVF Theorem.
\item By introducing sources for appropriate BRST transformations, and
integrating out half of the symplectic phase space variables, we can
formally derive a superfield analog of the field-antifield formalism.
\end{itemize}

\noi
The formalism we have set up seems to leave little room for an alternative
formulation, but we have clearly by no means proved that
the present construction gives a unique superfield formulation of the
BRST quantization program. This holds in particular for the extension
to the Lagrangian (field-antifield) formalism, where it is easy to
imagine that more suitable schemes may be found. The singular nature of the
solution to the superfield Master Equation is a consequence one would
like to avoid. But it is not presently clear whether the obstacle is
of fundamental or just technical nature.

\noi
Since we have constructed the appropriately gauge fixed phase-space path
integral, one could also try to derive from it, following conventional
lines, a relativistically covariant Lagrangian formulation for specific
field theories. This would be of interest for comparison with known
earlier attempts at writing a superfield version of the Lagrangian path
integral for such gauge theories. 

\noi
Our construction has turned out to identify the superspace behind the 
quantization of theories with first-class constraints precisely as
one would have expected. Viewed as a two-dimensional space spanned
by $t$ and the fermionic coordinate $\theta$, evolution in the $t$-direction
is generated by the Hamiltonian as in eq. (\ref{sh1}), while evolution
in the $\theta$-direction is generated by the BRST operator, as in 
eq. (\ref{shomega}). This provides us with a nice geometrical interpretation
of BRST symmetries. Anti-BRST symmetries can clearly
be understood in a similar manner, while the imposition of both BRST and 
anti-BRST symmetries simultaneously will require a new and enlarged framework.

\vspace{1cm}
\noindent
{\sc Acknowledgement:}~I.A.B.\ would like to thank the Niels Bohr Institute 
and Uppsala University for the warm hospitality extended to him there.
The work of I.A.B.\ and P.H.D.\ is partially supported by grant INTAS-RFBR
95-0829. 
The work of I.A.B.\ is also supported by grants INTAS 93-2058, INTAS 93-0633,
RFBR 96-01-00482, RFBR 96-02-17314, and NorFA 97.40.002-O.
The work of K.B.\ is partially supported by Nordita.

\vspace{1cm}
\appendix

\setcounter{equation}{0}

\section{Alternative Derivation of the Superfield BFV Theorem}

\noi
As we mentioned in Section \ref{secgaugeindsuper}, 
it is possible to prove the superfield 
analog of the BFV Theorem by means of variations induced by $Q_{\Psi}$
rather than by the original BRST charge $\Omega_{\Psi_{0}}$ itself. Since
this perhaps is conceptually more satisfactory (in the superfield
formulation we have seen that neither the BRST charge $\Omega$ nor the
Hamiltonian $H$ play fundamental r\^{o}les; only their ``superfield''
combinations in terms of $Q$ and $\bar{Q}$ enter), we reproduce this
alternative derivation here. It is clear from \eq{qdef1}, that we have 
to restrict ourself to ``exponential'' gauges
\mb{\Psi=\Psi'=\Psi'_{0}+\theta\Psi'_{1}} with \mb{\Psi''=0} 
in order for $Q_{\Psi}$ to
take over $\Omega_{\Psi_{0}}$'s dominant r\^{o}le.
The basic ingredients in the proof are of
course the same, and the proof itself leads to precisely the same 
conclusions as in Section \ref{secgaugeindsuper}. 
This is the first place where explicitly
need both of the superspace derivatives (\ref{d}). 

\noi
We start again with the path integral expression
\beq
{\cal Z} ~=~ \int [d\zzzz] ~ \exp\left[\frac{i}{\hbar}S(\zzzz)\right] ~,
\label{appsusypathint}
\eeq
with the superfield action
\bea
S(\zzzz) &=& \int\! dt~d\theta~ \left[
\hf \zzzz^{A}(t,\theta)~\omega_{AB}~
D\zzzz^{B}(t,\theta)(-1)^{\epsilon_{B}} 
- Q_{\Psi}(\zzzz(t,\theta),\theta)\right] \cr
&=& - \int\! dt~d\theta~ \left[
\hf(D\zzzz^{A}(t,\theta))~ \omega_{AB}~\zzzz^{B}(t,\theta)
+ Q_{\Psi}(\zzzz(t,\theta),\theta)\right] ~,
\label{appsusyaction}
\eea

\noi
Consider now the combination of a generalized superfield BRST-like
transformation, and a transformation with a superderivative:
\beq
\delta\zzzz^A(t,\theta) ~=~ \mu ~\{Q_{\Psi}(t,\theta),\zzzz^A(t,\theta)\}
- \mu~ \bar{D}\zzzz^A(t,\theta) ~.
\eeq
The first part of this transformation is what one might have guessed should
have played the r\^{o}le of a superfield BRST transformation. In fact, this
is not quite correct, since it does not leave the first (``kinetic energy'')
part of the action (\ref{appsusyaction}) invariant. This is the origin of
the compensating second piece of the transformation, which involves one
of the superderivatives.

\noi
We choose the fermionic parameter $\mu$ to be
\beq
\mu ~=~ \frac{i}{2\hbar} \int\! dt ~d\theta~ \EE(\delta\Psi') ~.
\eeq
This transformation is precisely tailored to yield no change in the action 
(\ref{appsusyaction}). It does, however, induce a non-trivial Jacobian:
\bea
  J-1&=& (-1)^{\epsilon_{A}} \int \! \int \!  dt ~d\theta~ 
  \delta(t-t')~ \delta(\theta-\theta')~ dt' ~d\theta' ~
 \dedel{\zzzz^{A}(t',\theta')} \delta\zzzz^{A}(t,\theta) \cr
&=& {i \over {2\hbar}} \int \!  dt ~d\theta~ \left[
 (\EE(\delta\Psi'))_{1}(t,\theta)
- \{  \EE(\delta\Psi')(t,\theta),~ Q_{\Psi}(t,\theta) \} \right]~.
\label{contrib1}
\eea 
While the last term \mb{\{\EE(\delta\Psi'),Q_{\Psi}\}}, 
as we shall see shortly, could be absorbed into 
a modified superfield gauge boson, 
the first term \mb{\EE(\delta\Psi')_{1}}
would spoil the interpretation as just a change
in gauge fixing. This should not come as a surprise, as a change in
$\Psi$ in general involves a canonical transformation as well. Indeed,
let us next perform a superfield canonical transformation of the kind
\beq
\delta\zzzz^A(t,\theta)
 ~=~ {1 \over 2} \{\EE(\delta\Psi')(t,\theta),\zzzz^A(t,\theta)\} ~.
\eeq
Due to the two compensating components of the superfield,
the measure is left invariant. But from the action we get a change
\beq
\delta S ~=~ -\frac{1}{2}\int \! dt~
 d\theta~\left[\{\EE(\delta\Psi')(t,\theta),
Q_{\Psi}(t,\theta)\} + (\EE(\delta\Psi'))_{1}(t,\theta)\right] ~.
\label{contrib2}
\eeq
Collecting the variations \e{contrib1} and \e{contrib2},
we see that the two \mb{\EE(\delta\Psi')_{1}} terms cancel while
the terms \mb{\{\EE(\delta\Psi'),Q_{\Psi}\}} add up to produce 
a term in the action of the form
\beq
\delta S ~=~ -\int \! dt~d\theta~ 
\left\{\EE(\delta\Psi')(t,\theta),Q_{\Psi}(t,\theta)\right\} ~.
\eeq

\noi 
Finally one notices that this just corresponds to the variation of 
$Q_{\Psi}$ under the infinitesimal change \mb{\Psi' \to \Psi' + \delta\Psi'}:
\bea
 Q_{\Psi+\delta\Psi} &~=~& Q_{\Psi} + \left\{
\int_0^1 \! d\alpha~ e^{\alpha {\mbox{\rm ad}}\Psi_{0}}\delta\Psi',
Q_{\Psi}\right\} \cr
&~=~&  Q_{\Psi} + \{\EE(\delta\Psi'),Q_{\Psi}\} ~.
\eea
Thus \mb{{\cal Z}_{\Psi+\delta\Psi} = {\cal Z}_{\Psi}}.
This establishes the proof of gauge independence based entirely on
$Q_{\Psi}$, rather than on the BRST operator $\Omega_{\Psi_{0}}$.

\setcounter{equation}{0}

\section{Superconventions}

Our convention for the Berezin integration is 
\beq
\int d\theta \ 1 = 0~,~~~~~~~~~~~~~~~\int  d\theta \ \theta  = 1~.
\eeq
The delta function can conveniently be represented as 
\beq
\delta(\theta)~=~\theta ~.
\eeq
It satisfies
\beq
\int F(\theta')~ \delta(\theta-\theta')~ d\theta'~=~ F(\theta) 
~=~ \int d\theta'~ \delta(\theta'-\theta)~ F(\theta')~.
\eeq
A superfield \mb{\zzzz(\theta) = \zzzz_{0} +\theta \zzzz_{1}} has
a functional derivative of opposite statistics 
\mb{\epsilon(\dede{\zzzz(\theta)})=\epsilon(\zzzz)+1}:
\beq
\dedel{\zzzz(\theta)} = \theta \dedel{\zzzz_{0}} 
 +(-1)^{\epsilon(\zzzz)}   \dedel{\zzzz_{1}}~,~~~~~~~~
\deder{\zzzz(\theta)} = -\deder{\zzzz_{0}}  \theta +  \deder{\zzzz_{1}}~,
\eeq
where $z_0$ and $z_1$ are independent variables. This means that
\beq
\dedel{\zzzz(\theta)} \zzzz(\theta') ~=~  \delta(\theta-\theta') 
~=~ \zzzz(\theta) \deder{\zzzz(\theta')}~.
\eeq
Right and left derivatives are connected via the formula
\beq
F  \deder{\zzzz(\theta)} =  (-1)^{\epsilon(F)(\epsilon(\zzzz)+1)+1}
 \dedel{\zzzz(\theta)} F~,~~~~~~~~~~~~~~
F  \deder{\zzzz(\theta)} d\theta
 =  (-1)^{\epsilon(\zzzz)(\epsilon(F)+1)}d\theta 
\dedel{\zzzz(\theta)} F~.
\eeq
Let us also finally mention the chain rule:
\bea
\dedel{\zzzz^{A}(\theta)} F
&=&  \int  (  \dedel{\zzzz^{A}(\theta)} \zzzz^{'B}(\theta') ) 
\  d\theta'  \ ( \dedel{\zzzz^{'B}(\theta')} F)~, \cr
F\deder{\zzzz^{A}(\theta)} 
&=&  \int (F \deder{\zzzz^{'B}(\theta')}) \  d\theta'  \
( \zzzz^{'B}(\theta')  \deder{\zzzz^{A}(\theta)} ) ~.\label{chainrule}
\eea


\begin{thebibliography}{999}
\bibitem{BFV}E.S.~Fradkin and G.A.~Vilkovisky, Phys.~Lett. \ {\bf B55}
(1975) 224.\newline
I.A.~Batalin and G.A.~Vilkovisky, Phys.~Lett. \ {\bf B69} (1977) 309. \newline
E.S.~Fradkin and E.S.~Fradkina, Phys.~Lett. \ {\bf B72} (1978) 343. \newline
I.A.~Batalin and E.S.~Fradkin, Phys.~Lett. \ {\bf B122} (1983) 157.
\bibitem{BF0}I.A.~Batalin and E.S.~Fradkin, Ann.~Inst.\ Henri Poincar\'{e}, 
{\bf 49} (1988) 145.
\bibitem{BF}I.A.~Batalin and E.S.~Fradkin, Riv.~Nuovo Cim.\ {\bf 9} (1986) 1.
\bibitem{BRST}C.~Becchi, A.~Rouet and R.~Stora, Comm.~Math.~Phys.\
{\bf 42} (1975) 127.\newline
I.V.~Tyutin, Lebedev preprint FIAN 39 (1975).
\bibitem{FOS}S.~Ferrara, O.~Piquet and M.~Schweda, Nucl.~Phys.\ {\bf 119}
(1977) 493.\newline
K.~Fujikawa, Prog.~Theor.~Phys.\ {\bf 59} (1978) 2045.\newline
F.~Ore and P.~van Nieuwenhuizen, Nucl.~Phys.\ {\bf B204} (1982) 317.\newline
L.~Bonora, P.~Pasti and M.~Tonin, Ann.~Phys.\ (N.Y.) {\bf 144} (1982) 15.
\bibitem{HSV}C.M.~Hull, B.~Spence and J.L.~V\'{a}squez-Bello, Nucl.~Phys.\
{\bf B348} (1991) 108. 
\bibitem{KO}T.~Kugo and I.~Ojima, Prog.~Theor.~Phys.~Suppl.\ {\bf 66}
(1979) 1.
\bibitem{BV}I.A.~Batalin and G.A.~Vilkovisky, Phys.~Lett.\ {\bf 102B}
(1981) 27; Phys.~Rev.\ {\bf D28} (1983) 2567 [E: {\bf D30} (1984) 508];
Nucl.~Phys.\ {\bf B234} (1984) 106.
\bibitem{KN}O.M.~Khudaverdian and A.P.~Nersessian, Mod.~Phys.~Lett.\
{\bf A8} (1993) 2377.
\bibitem{nersessian} A.~Nersessian, 
{\em Equivariant Localization: BV-geometry and Supersymmetric Dynamics},
hep-th/9310013.
\bibitem{superBV}N.R.F.~Braga and A.~Das, Nucl.~Phys.\ {\bf B442} (1995)
655.\newline N.R.F.~Braga and S.M.~de Souza, Phys.~Rev.\ {\bf D53}
(1996) 916. \newline P.M.~Lavrov, P.Yu.~Moshin and A.A.~Reshetnyak,
Mod.~Phys.~Lett.\ {\bf A10} (1995) 2687.
\newline P.M.~Lavrov, Phys.~Lett.\ {\bf B366} (1996) 160. 
\newline E.M.C.~Abreu and N.R.F.~Braga, Phys.~Rev.\ {\bf D54} (1996) 4080.
\bibitem{bering} K.~Bering, hep-th/9709003.
\bibitem{GGT}G.V.~Grigoryan, R.P.~Grigoryan and I.V.~Tyutin, Sov. J. Nucl.
Phys. {\bf 53} (1991) 1058.
\bibitem{BT1}I.A. Batalin and I.V. Tyutin, Int. J. Mod. Phys. {\bf A8}
(1993) 2333.













\end{thebibliography}
 \end{document}